\documentclass[a4paper,11pt]{article}
\pdfoutput=1 

\usepackage{jheppub}
\usepackage{amsmath}
\usepackage{amssymb}
\usepackage{epsfig}
\usepackage{subfigure}
\usepackage{cancel}

\usepackage{color}
\usepackage{graphicx}
\usepackage[usenames,dvipsnames,table]{xcolor}
\usepackage{varioref}

\usepackage{paralist}

\newcommand{\Zp}{\ensuremath{Z^\prime} }
\newcommand{\Wp}{\ensuremath{W^\prime} }

\newcommand{\SU}[1]{\ensuremath{\mathrm{SU}(#1)}}
\newcommand{\U}[1]{\ensuremath{\mathrm{U}(#1)}}

\renewcommand{\eqref}[1]{Eq.~(\ref{#1})}

\def\nbar{\bar{N}}
\def\bbar{\bar{b}}

\def\d{{\rm d}}
\def\eps{\epsilon}

\def\lr{\left( }
\def\rr{\right) }

\def\bsp#1\esp{\begin{split}#1\end{split}}

\def\beq{\begin{equation}}
\def\eeq{\end{equation}}
\def\bea{\begin{eqnarray}}
\def\eea{\end{eqnarray}}

\title{NLO+NLL limits on $W'$ and $Z'$ gauge boson masses in general extensions of the Standard
 Model}

\author[a]{Tom\'{a}\v{s} Je\v{z}o,}
\author[b]{Michael Klasen,}
\author[b]{David~R.~Lamprea,}
\author[c]{Florian Lyonnet,}
\author[c]{Ingo Schienbein}

\affiliation[a]{Department of Mathematical Sciences, University of Liverpool,
 Liverpool L69 3BX, United Kingdom}

 \affiliation[b]{Institut f\"ur Theoretische Physik, Westf\"alische
 Wilhelms-Universit\"at M\"unster, Wilhelm-Klemm-Stra\ss{}e 9, D-48149
 M\"unster, Germany}

\affiliation[c]{Laboratoire de Physique Subatomique et de Cosmologie,
 Universit\'e Joseph Fourier/CNRS-IN2P3/ INPG,
 53 Avenue des Martyrs, F-38026 Grenoble, France}

\emailAdd{jezo@lpsc.in2p3.fr}
\emailAdd{michael.klasen@uni-muenster.de}
\emailAdd{david.lamprea@uni-muenster.de}
\emailAdd{florian.lyonnet@lpsc.in2p3.fr}
\emailAdd{schien@lpsc.in2p3.fr}

\abstract{QCD resummation predictions for the production of charged ($W'$) and neutral
($Z'$) heavy gauge bosons decaying leptonically are presented. The results of our
resummation code at next-to-leading order and next-to-leading logarithmic (NLO+NLL) accuracy
are compared to Monte Carlo predictions obtained with PYTHIA at leading order (LO)
supplemented with parton showers (PS) and FEWZ
at NLO and next-to-next-to-leading order (NNLO) for the $p_T$-differential and total cross
sections in the Sequential Standard Model (SSM) and general SU(2)$\times$SU(2)$\times$U(1)
models. The LO+PS Monte Carlo and NNLO fixed-order predictions are shown to agree
approximately with those at NLO+NLL at small and intermediate $p_T$, respectively, and the
importance of resummation for total cross sections is shown to increase with the gauge
boson mass. The theoretical uncertainties are estimated by variations of the
renormalisation/factorisation scales and of the parton densities, the former being significantly
reduced by the resummation procedure. New limits at NLO+NLL on $W'$ and $Z'$ boson masses
are obtained by reinterpreting the latest ATLAS and CMS results in general extensions of
the Standard Model.}

\keywords{$W'/Z'$ bosons, LHC, resummation}

\preprint{LPSC-14-220, LTH 1015, MS-TP-14-23}

\begin{document} 
\maketitle
\flushbottom

\section{Introduction}
\label{sec:1}

New charged and neutral resonances are predicted in many well-motivated extensions of
the Standard Model (SM) such as Grand Unified Theories (GUTs) or models with extra spatial
dimensions \cite{Beringer:1900zz}. These extensions generally do not predict the precise
energy scale, at which the new heavy states should manifest themselves. However, for
various theoretical reasons (e.g.\ the hierarchy problem), new physics is expected to
appear at the TeV scale and is searched for at the Large Hadron Collider (LHC), which will
soon operate at centre-of-mass energies of $\sqrt{S}= 13$ TeV (LHC13) and 14 TeV (LHC14).

Experimental searches for $W'$ and $Z'$ bosons have so far mostly been performed
in the Sequential Standard Model (SSM) \cite{Altarelli:1989ff} (see Tab.\
\ref{tab:1}), where identical couplings for the new and SM gauge bosons are assumed
and which thus serves as a benchmark for comparisons among different experiments, but
is theoretically unmotivated. While we also adopt this model as a baseline to
compare predictions with different theoretical accuracy, we then enlarge our analysis to a
general $\text{G}(221) \equiv \text{SU}(2)_{1} \times \text{SU}(2)_{2}\times \text{U}(1)_{X}$
gauge group, which represents a well-motivated intermediate step towards the unification
of the SM gauge groups. In this framework, constraints on the parameter space from
low-energy precision observables have been derived \cite{Hsieh:2010zr}, and several
aspects of the collider phenomenology have already been studied \cite{Jezo:2012rm,%
Cao:2012ng,Abe:2012fb,Jinaru:2013eya}. Furthermore, the effect of the new spin-one
resonances on the interactions of ultra-high energy neutrinos in the atmosphere has been
analysed \cite{Jezo:2014kla}. Several well-known models emerge naturally from different
ways of breaking the $\text{G}(221)$ symmetry down to the SM gauge group
\cite{Hsieh:2010zr}, in particular Left-Right (LR) \cite{Mohapatra:1974gc,%
Mohapatra:1974hk,Mohapatra:1980yp}, Un-Unified (UU) \cite{Georgi:1989ic,Georgi:1989xz}, 
Non-Universal (NU) \cite{Malkawi:1996fs,Li:1981nk}, Lepto-Phobic (LP), Hadro-Phobic
(HP), and Fermio-Phobic (FP) \cite{Barger:1980ix,Barger:1980ti} models.

The ATLAS and CMS collaborations have performed extensive searches of new spin-one
resonances at the LHC for a large number of final states. In Tab.\ \ref{tab:1},
%
\begin{table}[t]
\centering
\resizebox{\columnwidth}{!}{%
\begin{tabular}{lccccc}
\hline\hline
Reference & $\sqrt{S}$ [TeV] & ${\cal L}$ [fb$^{-1}$] & Mode & Limits [TeV]& Comments \\ [0.5ex]
\hline\hline
{\bf ATLAS:}
\\
PLB701(2011)50 \cite{Aad:2011fe} & 7 & 0.036 & $W' \to \ell \nu$ & $M_{W'} > 1.49$ & SSM
\\
PLB705(2011)28 \cite{Aad:2011yg} & 7 & 1.04 & $W' \to \ell \nu$ & $M_{W'} > 2.15$ & SSM
\\
EPJC72(2012)2241 \cite{Aad:2012dm} & 7 & 4.7 & $W' \to \ell \nu$ & $M_{W'} > 2.55$ & SSM
\\
ATLAS-CONF-2014-017 \cite{ATLAS:2014fk} & 8 & 20.3 & $W' \to \ell \nu$ & $M_{W'} > \mathbf{3.27}$ & SSM
\\
JHEP09(2014)037 \cite{ATLAS:2014wra} & 8 & 20.3 & $W'\to\ell\nu$ & $M_{W'}>3.24$ & SSM
\\
PRD85(2012)112012 \cite{Aad:2012vs} & 7 & 1.02 & $W' \to WZ \to \ell \nu \ell' \ell'$ & $\sigma \times$Br &
\\
PRL109(2012)081801\cite{Aad:2012ej} & 7 & 1.04 & $W'\to t b\to\ell\nu jj$ & $M_{W'_R} > 1.13$ & LR Model
\\
EPJC72(2012)2056  \cite{ATLAS:2012ak} & 7 & 2.1 & $W'_R \to \ell N \to \ell \ell j j$ & $(M_{W'_R}, M_{N})$ exclusions & LR Model
\\
PRD87(2013)112006 \cite{Aad:2013wxa} & 7 & 4.7 & $W'\to WZ \to \ell \nu jj$ & $M_{W' }> 0.95$ & 
\\
JHEP01(2013)29 \cite{ATLAS:2012pu} & 7 & 4.8 & $W'\to jj$  & $M_{W'} >  1.68$  & 
\\
ATLAS-CONF-2013-050 \cite{ATLAS-CONF-2013-050} & 8 & 14 & $W'\to t b \to \ell \nu b b$ & $M_{W'_L} >  1.74$, $M_{W'_R} > 1.84$ & LR Model
\\
CERN-PH-EP-2014-147 \cite{Aad:2014aqa} & 8 & 20.3 & $W'\to jj$ & $M_{W'}>2.45$ & SSM
\\
PLB737(2014)223 \cite{Aad:2014pha} & 8 & 20.3 & $W'\to WZ \to \ell \nu\ell'\ell'$ & $M_{W'} > 1.52$ & 
\\
CERN-PH-EP-2014-152 \cite{Aad:2014xra} & 8 & 20.3 & $W'\to t b \to qq b b$ & $M_{W'_L} >  1.68$, $M_{W'_R} > 1.76$ & LR Model
\\
\hline
PLB700(2011)163 \cite{Aad:2011xp} & 7 & 0.04 & $Z'\to \ell \ell$ & $M_{Z'} >  1.048$ & SSM
\\
PRL107(2011)272002 \cite{Collaboration:2011dca} & 7 & 1.08-1.21 & $Z'\to \ell \ell$ & $M_{Z'} >  1.83$ & SSM
\\
JHEP11(2012)138 \cite{Aad:2012hf} & 7 & 4.9 & $Z'\to \ell \ell$ & $M_{Z'} >  2.22$ & SSM
\\
CERN-PH-EP-2014-053 \cite{Aad:2014cka} & 8 & 20.3-20.5 & $Z'\to \ell \ell$ & $M_{Z'} >  \mathbf{2.90}$ & SSM
\\
EPJC72(2012)2083 \cite{Aad:2012wm} & 7 & 2.05 & $Z'\to t t$ & $\sigma \times$Br  & 
\\
PRD87(2013)052002 \cite{Aad:2012xsa} & 7 & 4.6 & $\ell \ell \ell$  & $\sigma^{\rm vis.}$  &
\\
PLB719(2013)242 \cite{Aad:2012gm} & 7 & 4.6 & $Z'\to \tau \tau$ & $M_{Z'} > 1.4$ & SSM
\\
PRD88(2013)012004 \cite{Aad:2013nca} & 7 & 4.7 & $Z'\to t t$ & $\sigma \times$Br  & Narrow $Z'$
\\
JHEP01(2013)116 \cite{Aad:2012raa} & 7 & 4.7 & $Z'\to t t $ & $\sigma \times$Br & 
\\
ATLAS-CONF-2013-052 \cite{TheATLAScollaboration:2013kha}& 8 & 14 & $Z'\to t t$ & $\sigma \times$Br & Narrow $Z'$
\\
ATLAS-CONF-2013-066 \cite{ATLAS-CONF-2013-066} & 8 & 19.5 & $Z'\to \tau \tau$ & $M_{Z'} > 1.9$ & SSM
\\
\hline\hline
{\bf CMS:}
\\
PLB698(2011)21 \cite{Khachatryan:2010fa} & 7 & 0.036 & $W' \to e \nu_e$ & $M_{W'} > 1.36$ & SSM
\\
PLB701(2011)160 \cite{Chatrchyan:2011dx} & 7 & 0.036 & $W' \to \mu \nu_\mu$ & $M_{W'} > 1.4$ & SSM
\\
JHEP08(2012)023 \cite{Chatrchyan:2012meb} & 7 & 5 & $W' \to \ell \nu$ & $M_{W'_L} > 2.43$-$2.63$, $M_{W'_R} > 2.5$ & LR Model
\\
PRD87(2013)072005 \cite{Chatrchyan:2013lga} & 7-8 & 5-3.7 & $W' \to \ell \nu$ & $M_{W'} > 2.9$ & SSM
\\
CERN-PH-EP-2014-176 \cite{Khachatryan:2014tva} & 8 & 19.7 & $W' \to \ell \nu$ & $M_{W'} > \mathbf{3.28}$ & SSM
\\
PLB704(2011)123 \cite{Chatrchyan:2011ns} & 7 & 1 & $W' \to jj$  & $M_{W'} >1.51 $ & SSM
\\
PRL109(2012)261802 \cite{CMS:2012zv}   & 7 & 5 & $W'_R \to \ell N$ & $(M_{W'_R}, M_{N})$ exclusions & LR Model
\\
PRL109(2012)141801 \cite{Chatrchyan:2012kk} & 7 & 5 & $W' \to WZ \to 3\ell \nu$  &  $M_{W'} > 1.143$  & SSM 
\\
JHEP02(2013)036 \cite{Chatrchyan:2012rva} & 7 & 5 & $W' \to WZ \to \ell\ell jj$  &  $M_{W'} >0.94$  & SSM 
\\
PLB723(2013)280 \cite{Chatrchyan:2012ypy} & 7 & 5 & $W' \to WZ \to 4j$  & $\sigma \times$Br   & SSM 
\\
PLB718(2013)1229 \cite{Chatrchyan:2012gqa} & 7 & 5 & $W' \to t b \to \ell \nu b b$ & $M_{W'_L} > 1.51$, $M_{W'_R} > 1.85$ & LR Model
\\
PLB717(2012)351 \cite{Chatrchyan:2012su} & 7 & 5 & $W' \to ttj$ & $M_{W'_R} > 0.84$  & LR Model
\\
CMS-PAS-EXO-12-025 \cite{CMS-PAS-EXO-12-025}& 8 & 19.5 & $W' \to WZ$  & $M_{W'} >1.47$  & SSM
\\
CERN-PH-EP-2014-161 \cite{Khachatryan:2014dka}& 8 & 19.7 & $W'_R \to \ell N$ & $(M_{W'_R}, M_{N})$ exclusions & LR Model
\\
JHEP08(2014)173 \cite{Khachatryan:2014hpa} & 8 & 19.7 & $W'\to WZ\to jjX$ & $M_{W'}>1.7$ & SSM \\
\hline
JHEP05(2011)093 \cite{Chatrchyan:2011wq} & 7 & 0.04 & $Z'\to \ell \ell$  &  $M_{Z'} > 1.14$ & SSM
\\
PLB714(2012)158 \cite{Chatrchyan:2012it} & 7 & 5 & $Z'\to \ell \ell$  &  $M_{Z'} > 2.33$ & SSM
\\
PLB720(2013)63 \cite{Chatrchyan:2012oaa} & 7-8 & 5.3-4.1 & $Z'\to \ell \ell$  &  $M_{Z'} > 2.59$ & SSM
\\
CMS-PAS-EXO-12-061 \cite{CMS:2013qca} & 8 & 19.6-20.6 & $Z'\to \ell \ell$  &  $M_{Z'} > \mathbf{2.96}$ & SSM
\\
PLB716(2012)82 \cite{Chatrchyan:2012hd} & 7 & 4.9 & $Z'\to \tau \tau$  &  $M_{Z'} > 1.4$ & SSM
\\
JHEP09(2012)029 \cite{Chatrchyan:2012ku} & 7 & 5 & $Z' \to t t $ &  $\sigma \times$Br  & 
\\
JHEP01(2013)013 \cite{CMS:2012yf} & 7 & 5 & $Z', W' \to jjX$, $Z' \to b b$  & $M_{W'}>1.92$, $M_{Z'} > 1.47$  & SSM
\\
PRD87(2013)114015 \cite{Chatrchyan:2013qha} & 8 & 4 & $Z',W' \to jj$  & $M_{W'}  >1.73 $, $M_{Z'} >1.62$  & 
\\
CMS-PAS-EXO-12-059 \cite{CMS-PAS-EXO-12-059}& 8 & 19.6 & $Z',W' \to jj$  & $M_{W'}  >2.29 $, $M_{Z'} >1.68$  & SSM
\\
CMS-PAS-EXO-12-023 \cite{CMS-PAS-EXO-12-023}& 8 & 19.6 & $Z' \to b b$  & $M_{Z'} >1.68$  & SSM
\\
\hline\hline
\end{tabular}
}
\caption{ATLAS and CMS searches for new spin-one gauge bosons ($W'$ and $Z'$) at the LHC
 using data from the $pp$ runs in 2010 and 2011 at $\sqrt{S}= 7$ TeV and from the $pp$ run
 in 2012 at $\sqrt{S}= 8$ TeV.}
\label{tab:1}
\end{table}
%
we summarise these searches, that exploited data from the $pp$ runs in 2010 and 2011
at $\sqrt{S}= 7$ TeV (LHC7) and from the $pp$ run in 2012 at $\sqrt{S}= 8$ TeV (LHC8),
as well as the corresponding constraints on $W'$ and $Z'$ gauge boson masses.
As can be seen, the most stringent limits come from searches with purely leptonic final
states, $W' \to \ell \nu$ \cite{Aad:2011fe,Aad:2011yg,Aad:2012dm,ATLAS:2014fk,%
ATLAS:2014wra,Khachatryan:2010fa,Chatrchyan:2011dx,Chatrchyan:2012meb,Chatrchyan:2013lga,%
Khachatryan:2014tva} and $Z' \to \ell \ell$ \cite{Aad:2011xp,Collaboration:2011dca,%
Aad:2012hf,Aad:2014cka,Chatrchyan:2011wq,Chatrchyan:2012it,Chatrchyan:2012oaa,CMS:2013qca}
(with $\ell = e, \mu$, neutrino flavours and antiparticles understood), leading to
(preliminary) lower mass limits of $M_{W'} \gtrsim 3.3$ TeV
\cite{ATLAS:2014fk,Khachatryan:2014tva} and $M_{Z'} \gtrsim 2.9$ TeV \cite{Aad:2014cka,%
CMS:2013qca} for gauge bosons in the SSM. In LR models,
exclusion contours in the right-handed weak boson ($W_R$) and neutrino ($N$) mass plane
have been obtained by exploiting also semileptonic \cite{Aad:2012ej,ATLAS:2012ak,%
ATLAS-CONF-2013-050,CMS:2012zv,Chatrchyan:2012gqa,Chatrchyan:2012su,Khachatryan:2014dka}
and even fully hadronic final states \cite{Aad:2014xra}. In addition, upper limits on the
production
cross section times the branching ratio, $\sigma \times{\rm Br}$, were presented, which
can be used to constrain a few other specific models such as extended gauge models with
modified couplings of the new to the SM gauge bosons \cite{Aad:2012vs,Aad:2012wm,%
Aad:2013nca,Aad:2012raa,TheATLAScollaboration:2013kha,Chatrchyan:2012ypy,Chatrchyan:2012ku}.
However, other G(221) models such as UU and NU models have so far not been analysed.
Furthermore, the mass limits are mostly obtained using LO+PS Monte Carlo simulations
with PYTHIA \cite{Sjostrand:2006za} rescaled to NNLO with FEWZ \cite{Gavin:2010az,%
Gavin:2012sy}, where both programs do a priori not include the important interference
effects of new and SM gauge boson exchanges.

In this paper, we present new QCD resummation predictions, which include these
interference effects, at next-to-leading order and next-to-leading logarithmic (NLO+NLL)
accuracy for the production of charged and neutral heavy gauge boson ($W'$ and $Z'$)
decaying into charged leptons and neutrinos. 
For SM weak gauge boson production, the importance of resummation calculations has been
demonstrated most recently using Soft-Collinear Effective Theory (SCET) by an improved
agreement with Tevatron and LHC data and reduced theoretical scale uncertainties
\cite{Becher:2011fc,Becher:2012xr,Becher:2013vva}.
In the context of new physics searches at the LHC, soft-gluon resummation
has already been applied to the production of $Z'$ bosons \cite{Fuks:2007gk}
as well as to the production of supersymmetric (SUSY) particles such as squarks and
gluinos \cite{Kramer:2012bx}, sleptons \cite{Bozzi:2006fw,Bozzi:2007qr,Bozzi:2007tea,%
Fuks:2013lya}, and gauginos \cite{Debove:2009ia,Debove:2010kf,%
Debove:2011xj,Fuks:2012qx}. For the $Z'$ boson and weak SUSY channels, the NLO+NLL code
RESUMMINO is publicly available \cite{Fuks:2013vua}. We have now also added the
possibility to make predictions for $W'$ bosons with general gauge couplings for
transverse momentum ($p_T$) spectra, resummed as $p_T\to0$ in impact parameter
space, and for total cross sections, resummed near partonic threshold in Mellin
space.\footnote{Our code is available at \texttt{http://www.resummino.org}.}

The results of our resummation code are compared using different benchmark models
to $p_T$ distributions and total cross sections obtained with the LO+PS Monte Carlo
generator PYTHIA \cite{Sjostrand:2006za}, in which we have implemented the new
weak bosons including the interferences with the SM gauge bosons. In addition, we
compare with the theoretical predictions in fixed order perturbation theory at NLO
and NNLO QCD calculated with the FEWZ program, which unfortunately lacks
the interference terms \cite{Gavin:2010az,Gavin:2012sy}. The theoretical uncertainties
are estimated by variations of the renormalisation/factorisation scales and parton
distribution functions (PDFs). Using the example of $Z'$ bosons, we demonstrate that
the resummation contributions become more important with increasing mass of the new
gauge boson, which will further increase their importance in future LHC analyses.
In an exemplary way, we reinterpret the most recent ATLAS $W'$ \cite{ATLAS:2014fk}
and CMS $Z'$ \cite{CMS:2013qca} analyses using our NLO+NLL predictions, including
interference, and three different new physics models.

The remainder of this paper is organised as follows: In Sec.\ \ref{sec:2} we describe
the relevant features of the three theoretical approaches (PYTHIA, FEWZ, and RESUMMINO)
that we develop, employ and compare in this paper. Theoretical numerical predictions,
i.e.\ differential and total cross sections for the production of charged and neutral
heavy gauge bosons decaying to leptons at the LHC and their associated theoretical
uncertainties, are presented in Sec.\ \ref{sec:3}. The reanalyses of the ATLAS and
CMS experimental results are described in Sec.\ \ref{sec:4}. Finally, we summarise
our results and draw our conclusions in Sec.\ \ref{sec:5}.

\section{Theoretical setup}
\label{sec:2}

In this section, we describe the main features of the three different theoretical
frameworks (PYTHIA, FEWZ, and RESUMMINO) that we have (in particular in the cases
of PYTHIA and RESUMMINO) developed further and that we employ and compare numerically
in Secs.\ \ref{sec:3} and \ref{sec:4}.

\subsection{PYTHIA Monte Carlo at LO+PS}

Following common experimental practice, we first simulate the production of new
charged ($W'$) and neutral ($Z'$) gauge bosons decaying leptonically into $\ell \nu$
and $\ell\ell$ (with $\ell = e, \mu$, neutrino flavours and antiparticles understood)
at LO, ${\cal O}(\alpha^2\,\alpha_{s}^0$), using PYTHIA 6.4.27 \cite{Sjostrand:2006za}. The
description of kinematic distributions is improved  to leading-logarithmic accuracy
by a virtuality-ordered PS, applied in this case only to the initial partons and
introducing a weak dependence on the strong coupling constant $\alpha_s$ (or equivalently
the QCD scale $\Lambda$) even at this order. Still, we estimate the scale uncertainty
of the PYTHIA LO+PS prediction by varying only the factorisation scale $\mu_F$ by a
factor of two about the central value, set by the new gauge boson mass. We use PDFs
from the MSTW 2008 parameterisation, i.e.\ in this case the central fit MSTW 2008 LO
\cite{Martin:2009iq}.

In PYTHIA, the $Z'$ and $W'$ bosons (PDG codes 32 and 34) constitute hypothetical physical
(mass eigenstate) vector bosons. The $W'$ boson couples, e.g., to the SM fermions with 
strengths
\beq
 \bar{\nu}_{\ell}\ell W'^+, \bar{\ell} \nu_{\ell} W'^- \sim
 \frac{g}{2\sqrt{2}} \gamma^{\mu}(V_{\ell} - A_{\ell} \gamma_{5}) \ , \quad
 \bar{q}q'W'^\pm  \sim
 \frac{g}{2\sqrt{2}} U_{\rm CKM}\gamma^{\mu}(V_{q} - A_{q} \gamma_{5}) \ .
 \label{eqn:pythia_couplings}
\eeq
Here, $g$ is the SU(2) coupling constant related to the fine structure constant
$\alpha=g^2\sin^2\theta_W/$ $(4\pi)$ through the weak mixing angle $\theta_W$
and calculated numerically at the scale of the new gauge boson mass using
input values from the Particle Data Group \cite{Beringer:1900zz}. We thus
obtain $\alpha(M_Z)=1/128.97$ and $\alpha(M_{V'}=4$ TeV$)=1/123.36$.
$U_{\rm CKM}$ is the quark mixing matrix, and 
the couplings $V_{\ell,q}$ and $A_{\ell,q}$ are dimensionless, real, and fermion generation
independent. Their default values are $V_{\ell,q}=1$ and $A_{\ell,q}=-1$ as in the SSM.
Similarly, the $Z'$ boson couplings are set to their SSM values, but may be modified
by the user. It is also possible to allow additional couplings to SM vector and Higgs
bosons when necessary, e.g., for general extended gauge models.
The total decay widths $\Gamma_{V'}$ of the new vector bosons ($V'$)
are calculated perturbatively
in an automated fashion as a sum of the decay widths into SM fermions, taking into account
the user-provided values of $V_{\ell,q}$ and $A_{\ell,q}$. We have verified that in the
models that we will consider (SSM, UU and NU models) the decays into pairs
of gauge and Higgs bosons contribute only 1-2\% to the total decay widths, so that we
may safely neglect them. In other models there may of course be regions of parameter
space where these decays are not negligible \cite{Jinaru:2013eya}. In the Breit-Wigner
propagator, a centre-of-mass energy ($s$) dependence may furthermore be introduced
in the terms dependent on the total decay width, $M_{V'}\Gamma_{V'}\to s\Gamma_{V'}/M_{V'}$,
to improve the description of the resonance shape \cite{Berends:1989aa}.

Since the $2\to1\to2$ structure of the PYTHIA implementation of new vector boson
production and decay does not easily lend itself to taking into account interference effects and
since these can be quite important for kinematic distributions and also for
total cross sections depending on experimental cuts \cite{Rizzo:2007xs, Accomando:2013sfa},
we have implemented the full processes $qq^{(\prime)} \to \ell\ell\,(\ell \nu)$ including also
interferences of SM and new gauge bosons, but still neglecting the masses of the
final state leptons. Here, the couplings have been implemented both in a general fashion
(see above) and for specific G(221) models. This has also been done in the routines
calculating the total decay widths.\footnote{The modified PYTHIA code is available from
the authors upon request.} 

\subsection{Perturbative QCD at NLO and NNLO}

In fixed-order perturbative QCD, the hadronic production of $Z'$ and $W'$ bosons can
be calculated at NLO, ${\cal O}(\alpha^2\alpha_s$), and NNLO, ${\cal O}(\alpha^2\alpha_s^2$),
with the publicly available program FEWZ \cite{Gavin:2010az,Gavin:2012sy} in a fully
exclusive way and including the leptonic decay of the gauge bosons with full spin
correlations and finite width effects. FEWZ thus allows one to investigate the total production
cross section as well as the transverse momentum and invariant/transverse mass spectra
under arbitrary kinematic cuts on the gauge boson and/or lepton-pair. Unfortunately,
since the code assumes the existence of one neutral/charged vector boson only, the
important interference effects between two different gauge bosons are not taken into
account. The pure SM background is, however, far below the new physics signal, at least
for the new gauge boson masses and models considered here, and can thus safely be neglected.

Since the user is allowed to tune the gauge boson properties such as the mass, width and
partial width into leptonic states, FEWZ can in principle be used to extrapolate SM
predictions to the production rate of $Z'$ and $W'$ bosons in various extensions of the SM.
Unfortunately, it is not possible to enter the gauge boson couplings to SM fermions directly,
so that additional rescaling of the observables may be required.
In order to calculate the observables of the heavy resonance in a given extension of the SM,
we proceed therefore in three steps:
\begin{inparaenum}[(i)]
 \item we obtain the $Z'$ and $W'$ boson properties from our extended version of PYTHIA,
 discussed in the previous section, and feed them into FEWZ;
 \item we calculate the total cross section and desired distributions of a $Z'$ or $W'$
 resonance with SM couplings;
 \item we rescale the observables by the relevant combination of $Z'$ and $W'$ boson
 couplings to SM fermions.
\end{inparaenum}
This rescaling of the cross section must be done carefully and is in certain models impossible.
The squared matrix element calculated in FEWZ is given by \cite{Melnikov:2006kv}
\beq
 \left|\mathcal{M}\right|^{2} =
 \frac{H^{\mu\nu}L_{\mu\nu}}{\left( Q^{2}-M_{V'}^{2} \right)^{2} + M_{V'}^2\Gamma_{V'}^2}\,,
 \label{eq:FEWZmatrixelement}
\eeq
where $Q^2$ is the invariant mass of the di-lepton pair, $H^{\mu\nu}$ is the hadronic tensor
including the QCD corrections, and $L_{\mu\nu}$ is the leptonic tensor. For ease of use,
$L_{\mu\nu}$ is expressed in terms of $M_{V'}$ and $\mathrm{Br}(Z',W'\rightarrow \ell\ell,\ell
\nu)$ rather than in terms of the corresponding couplings to leptons. Consequently, when
considering $Z'$ and $W'$ resonances, only their couplings to the initial quarks must be
rescaled. We estimate the scale uncertainty of the FEWZ prediction by varying the
renormalisation scale $\mu_R$ and factorisation scale $\mu_F$ simultaneously by a factor
of two about the central value, the new gauge boson mass. The central PDF parameterisations
are MSTW 2008 NLO and NNLO, respectively, and their uncertainties are estimated with
the 40 sets of error PDFs at 68\% confidence level as implemented in the FEWZ code
\cite{Martin:2009iq}.

\subsection{Resummation at NLO and NLL}

In this section, we briefly review the formalism that allows us to
resum the QCD corrections to all orders at large invariant mass
($Q$) and/or small transverse momentum ($p_T$) of a lepton pair
produced through weak gauge bosons in hadronic collisions \cite{Fuks:2013vua}.

Thanks to the QCD factorisation theorem, the double differential cross section
\bea
  Q^2\frac{\d^2\sigma_{AB}}{\d Q^2 \d p_T^2}(\tau)&=&\sum_{ab}
  \int_0^1 \!\d x_a \d x_b \d z[x_a f_{a/A}(x_a,\mu_F^2)] [x_b f_{b/B}(x_b,
  \mu_F^2)]\,[z\,\d\sigma_{ab}(z,Q^2,p_T^2,\mu_F^2)]\nonumber\\
 &\times& \delta(\tau-x_ax_bz)
  \label{eq:HadFacX}
\eea
can be obtained by convolving the partonic cross section $\d\sigma_{ab}$ with
the universal densities $f_{a,b/A,B}$ of the partons $a,b$, carrying the momentum
fractions $x_{a,b}$ of the colliding hadrons $A,B$, at the factorisation scale
$\mu_F$. The application of a Mellin transform
\bea
  F(N)&=&\int_0^1 \d y \,y^{N-1} F(y)
  \label{eq:MelDef}
\eea
to the quantities $F\in\{\sigma_{AB},~\sigma_{ab},~f_{a/A},~f_{b/B}\}$ with
$y\in\{\tau=Q^2/S,~z=Q^2/s,~x_{a},~x_{b}\}$ allows to express the hadronic cross
section in moment space as a simple product,
\bea
  Q^2\frac{\d^2\sigma_{AB}}{\d Q^2 \d p_T^2}(N-1)&=&\sum_{ab} f_{a/A}(N,\mu_F^2) 
  f_{b/B}(N,\mu_F^2) \sigma_{ab}(N,Q^2,p_T^2,\mu_F^2).
  \label{eq:HadFacN}
\eea
Furthermore, the application of a Fourier transform to the partonic cross section
$\sigma_{ab}$ allows to correctly take into account transverse-momentum
conservation, so that in moment ($N$) and impact parameter ($b$) space it can be
written as
\bea
 \sigma_{ab}(N,Q^2,p_T^2,\mu_F^2)&=&
 \int_0^\infty \d b\frac{b}{2} J_0(bp_T) \sigma_{ab}(N,Q^2,b^2,\mu_F^2).
 \label{eq:JR:pff}
\eea
Here, $J_0(y)$ denotes the $0^{\rm th}$-order Bessel function and
\bea
 \sigma_{ab}(N,Q^2,b^2,\mu_F^2)&=&\sum_{n=0}^\infty a_s^{n}(\mu_R^2)\,
 \sigma_{ab}^{(n)}(N,Q^2,b^2,\mu_F^2,\mu_R^2)
 \label{eq:5}
\eea
is usually expanded perturbatively in the strong coupling constant $a_s(\mu^2)=
\alpha_s(\mu^2)/(2\pi)$ at the renormalisation scale $\mu_R$. For simplicity,
we identify in the following the factorisation and renormalisation scales, {\em
i.e.} $\mu_F=\mu_R=\mu$.

In the Born approximation, the production of lepton pairs is induced by quarks
$q$ and antiquarks $\bar{q}'$ in the initial (anti-)protons and is mediated by
$s$-channel electroweak gauge-boson exchanges, whose mass and couplings determine
the partonic cross section $\sigma_{q\bar{q}'}^{(0)}$. At ${\cal O}(a_s)$, virtual loop and
real parton emission corrections must be taken into account. The latter induce not
only a deviation of the partonic
centre-of-mass energy $s$ from the squared invariant mass $Q^2$ of the lepton
pair, but also non-zero transverse momenta $p_T$, that extend typically to values
of the order of the weak gauge boson mass.
Close to the partonic production threshold, where $z=Q^2/s\to1$ or $N\to\infty$,
the convergence of the perturbative expansion is spoiled due to soft gluon
radiation, which induces large logarithms
\bea
 a_s^n\lr \ln^m(1-z)\over1-z\rr_+&\to&a_s^n\ln^{m+1}\nbar+\dots
\eea
with $m\leq2n-1$ and $\nbar=N e^{\gamma_E}$ \cite{Bozzi:2007qr,Debove:2010kf}. Similarly, in
the small-$p_T$ (or large-$b$) region, where the bulk of the events is produced,
the convergence of the perturbative expansion is again spoiled by soft gluon
radiation, which induces large logarithms
\bea
 \alpha_s^n\lr{1\over p_T^2}\ln^m{Q^2\over p_T^2}\rr_+&\to&
 \alpha_s^n\ln^{m+1}\bbar^2+\dots
\eea
with $m\leq2n-1$ and $\bbar=bQe^{\gamma_E}/2$ \cite{Bozzi:2006fw,Debove:2009ia}. An important
observation, first
made by Li \cite{Li:1998is} and then further developped by Laenen, Sterman, and
Vogelsang \cite{Laenen:2000de,Laenen:2000ij} is that the common kinematic origin
of these divergences allows for a joint resummation of the large logarithms in the
partonic cross section.
In the corresponding kinematic limits and with proper adjustments,
the jointly resummed cross section
reduces to the one for transverse-momentum \cite{Debove:2009ia}
and threshold resummation \cite{Debove:2010kf}, respectively.

While the large logarithms must clearly be resummed close to the production
threshold, when $z\to1$ and $\nbar\to\infty$, and/or at small values of $p_T\to0$,
when $\bbar\to\infty$, they account only
partially for the full perturbative cross section away from these regions. In
order to obtain a valid cross section at all values of $z$ and $p_T$, the
fixed-order (f.o.) and the resummed (res.) calculations must be combined
consistently by subtracting from their sum the perturbatively expanded (exp.)
resummed component,
\bea
 \sigma_{ab}&=&
 \sigma^{\rm(res.)}_{ab}+\sigma^{\rm(f.o.)}_{ab}-\sigma^{\rm(exp.)}_{ab}.
 \label{eq:JR:Mat}
\eea
The latter is easily obtained by expanding Eq.\ (\ref{eq:JR:pff}) to the desired
accuracy.

After the resummation of the partonic cross section has been performed in $N$-
and $b$-space, we have to multiply the resummed cross section and its perturbative
expansion with the moments of the PDFs $f_{a/A}(N,\mu^2)$ and transform the
hadronic cross section obtained in this way back to the physical $z$- and
$p_T$-spaces. The moments of the PDFs are obtained through a numerical fit to the
publicly available PDF parameterisations in $x$-space.

Our NLO fixed order and NLO+NLL
resummation calculations have been implemented in the computer code RESUMMINO
that is publicly available \cite{Fuks:2013vua}. For our numerical results at both NLO and
NLO+NLL, we employ the PDF parameterisation of MSTW 2008 NLO and estimate the
theoretical scale error by varying again simultaneously the renormalisation
and factorisation scales by a factor of two about the new gauge boson mass.

\section{Numerical results}
\label{sec:3}

In this section, we present numerical results using the three different theoretical
approaches discussed above. We first fix the SM input parameters \cite{Beringer:1900zz},
select three new physics models \cite{Altarelli:1989ff,Hsieh:2010zr}, impose constraints
on their parameter spaces from a previous global analysis \cite{Hsieh:2010zr}, and select
five specific benchmark points in these models. We then compare
the transverse momentum spectra of SSM $W'$ bosons in the three theoretical approaches,
finding approximate agreement in the relevant kinematic regions, and we also compute
the corresponding scale uncertainties. Total cross sections are then presented for all
five selected benchmark points within the three theoretical frameworks, without and
with interference effects, and including not only scale, but also PDF uncertainties.
Finally, we demonstrate, using the example of SSM $Z'$ bosons, that the importance of
resummation effects increases with the invariant mass of the decay lepton pair.

\subsection{Input parameters}

The numerical results in this section are computed for $pp$ collisions at the LHC
with a hadronic centre-of-mass energy of $\sqrt{S}=14$ TeV (LHC14). The PDFs are
taken from the MSTW 2008 global fits at LO, NLO and NNLO, respectively, and the
corresponding error sets at 68\% C.L., following the prescription in Eqs.\ (50)--(52) 
of Ref.\ \cite{Martin:2009iq}. The renormalisation and
factorisation scales $\mu_R$ and $\mu_F$ are identified with the new gauge boson mass
$M_{V'}$, varied by a common factor of two up and down to estimate the scale uncertainty.
At LO, the strong coupling constant influences only differential cross sections
calculated with the PYTHIA PS, so that the corresponding default value is retained.
Beyond this order, $\alpha_s$ enters directly and is adopted from the PDG value at
NLO and NLL, as are the electromagnetic fine structure constant $\alpha$ and the squared
sine of the weak mixing angle $\theta_W$ \cite{Beringer:1900zz}, and from the (almost
identical) MSTW 2008 global fit value at NNLO \cite{Martin:2009iq}. This information is
summarised in Tab.\ \ref{tab:2}.

%
\begin{table}[t]
 \center
 \begin{tabular}{llcccc}
 \hline\hline
 Theory        & PDFs       & $\mu_{R,F}$ & $\alpha_s(M_Z)$  & $\alpha(M_Z)$ & $\sin^2\theta_W$\\
 \hline\hline
 PYTHIA LO+PS  & MSTW 2008 LO  & $M_{V'}$ & 0.130 & $1/128.97$ & 0.23116 \\
 FEWZ NLO      & MSTW 2008 NLO & $M_{V'}$ & 0.118 & $1/128.97$ & 0.23116 \\
 FEWZ NNLO     & MSTW 2008 NNLO& $M_{V'}$ & 0.117 & $1/128.97$ & 0.23116 \\
 RESUMMINO LO  & MSTW 2008 LO  & $M_{V'}$ & 0.118 & $1/128.97$ & 0.23116 \\
 RESUMMINO NLO & MSTW 2008 NLO & $M_{V'}$ & 0.118 & $1/128.97$ & 0.23116 \\
 RESUMMINO NLO+NLL & MSTW 2008 NLO & $M_{V'}$ & 0.118 & $1/128.97$ & 0.23116 \\
 \hline\hline
 \end{tabular}
 \caption{PDF and scale choices as well as SM input parameters used in our
 different theoretical calculations.
 At LO, the strong coupling constant influences only differential cross sections
 calculated with the PYTHIA PS, so that the corresponding default value is retained.}
 \label{tab:2}
\end{table}
%

Apart from the SSM with identical fermion couplings of SM and new gauge
bosons \cite{Altarelli:1989ff}, we study also the so-called G(221) models
\cite{Hsieh:2010zr}, which are based on the intermediate semi-simple group
$\SU{2}_1 \times \SU{2}_2 \times \U{1}_X$ with gauge couplings $g_i$,
$i\in\{1,2,X\}$. They can be categorised in two classes:
\begin{inparaenum}[(i)]
 \item Models, in which the first $\SU{2}$ subgroup is identified with the
 $\SU{2}_L$ of the SM and one breaks $\SU{2}_2\times \U{1}_X\to \U{1}_Y$ at
 some high scale $u$ with Higgs doublets or triplets. They include in particular
 the LR model \cite{Mohapatra:1974gc,Mohapatra:1974hk,Mohapatra:1980yp},
 motivated by non-zero neutrino masses and the prospects of parity restoration
 and the existence of right-handed neutrinos and studied already in part by the
 ATLAS and CMS collaborations. They also include the LP, HP and FP models
 \cite{Barger:1980ix,Barger:1980ti}, irrelevant for the leptonic channels at
 the LHC studied here.
 \item Models, in which the $\U{1}$ subroup is identified with the $\U{1}_Y$
 of the SM and one breaks $\SU{2}_1\times\SU{2}_2\to\SU{2}_L$ with a Higgs
 bi-doublet, include the Un-Unified (UU) \cite{Georgi:1989ic,Georgi:1989xz}
 and Non-Universal (NU) \cite{Malkawi:1996fs,Li:1981nk} models. They are
 motivated by the large mass hierarchy of the SM fermions, in particular of
 quarks vs.\ leptons or of first and second generation vs.\ third generation
 fermions, and are accessible in leptonic channels at the LHC. In these
 models, $M_{Z'}^2/M_{W'}^2= 1+{\cal O}(v^2/u^2)$, where $v=246$ GeV is the vacuum
 expectation value (VEV) of the (SM-like) Higgs field of the second stage
 breaking and one assumes that the first stage Higgs VEV $u^2\gg v^2$.
\end{inparaenum}
Apart from $u$ or, equivalently, $M_{V'}=M_{W'}\simeq M_{Z'}$, their second free
parameter is the tangent of the mixing angle $\phi$ at the first breaking stage,
\beq
 t=\tan\phi={g_2\over g_1},
\eeq
to which the fermionic left-handed $Z'$ and $W'$ boson couplings are, modulo small
corrections of ${\cal O}(\eps\sim t/M_{V'}^2)$, proportional or anti-proportional.
This implies in particular for the rescaling of the FEWZ predictions that they
must be multiplied by a factor $t^2$ or $1/t^2$.

%
\begin{figure}[t]
 \begin{center}
 \includegraphics[width=\textwidth]{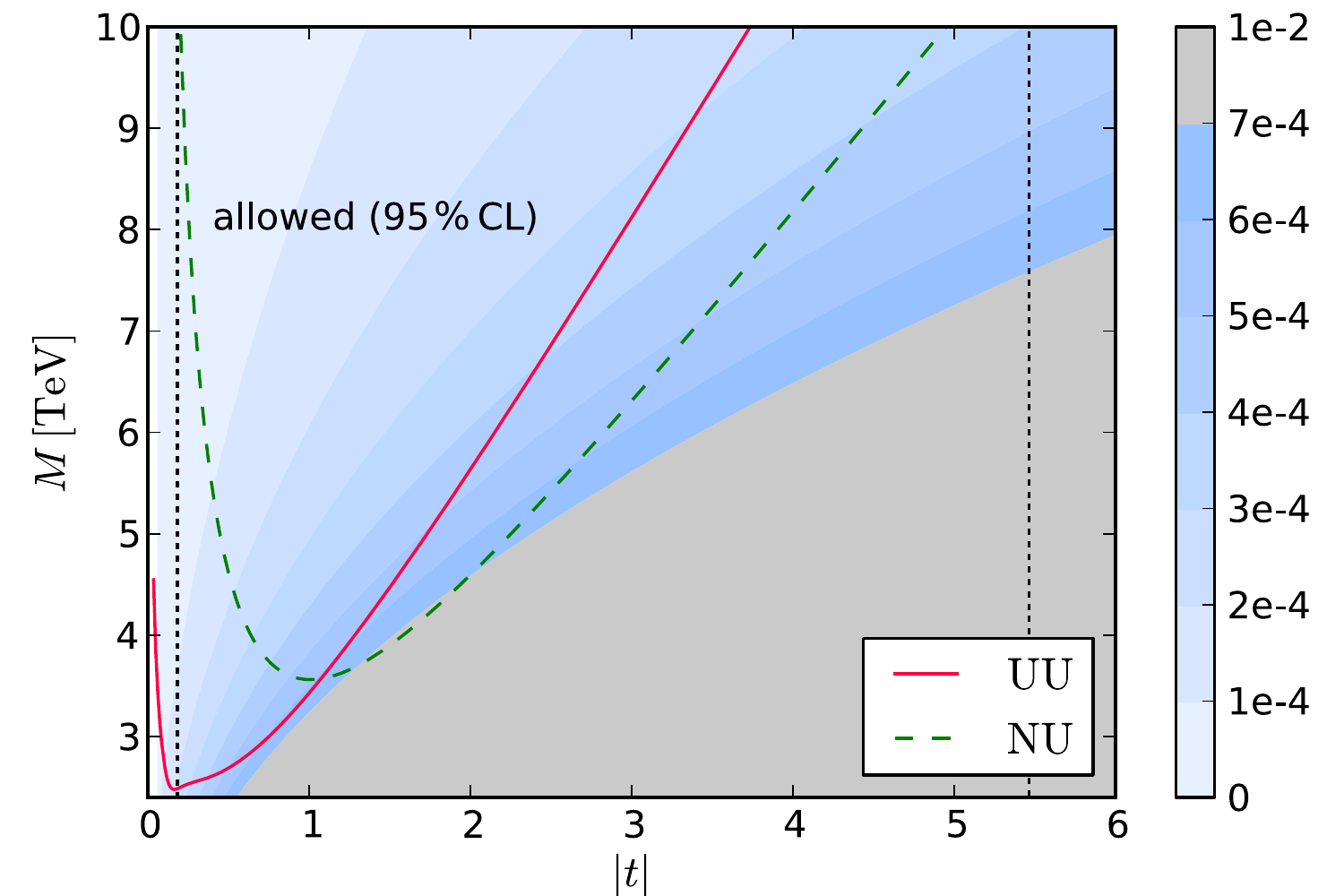}
 \caption{Exclusion limits for left-handed G(221) models. The red (full) and green (dashed)
 lines represent 95\% confidence level contours of allowed regions in the UU and NU models.
 In regions outside the area bounded by dotted lines at least one of the gauge couplings
 becomes non-perturbative. Shaded contours represent values of $\eps(t,M_{V'})$.}
 \label{fig:1}
 \end{center}
\end{figure}
%
In Fig.\ \ref{fig:1}, we have translated perturbativity ($g_i<\sqrt{4\pi}$) as
well as the low-energy and electroweak precision constraints obtained in Ref.\
\cite{Hsieh:2010zr} into allowed regions in the physical parameters
$t$ and $M_{V'}$. Coupling corrections of ${\cal O}(\eps)$ are indicated
as shaded bands and remain small in the allowed regions. As one can see, these
indirect constraints can be quite competitive compared to the direct LHC limits
(cf.\ Tab.\ \ref{tab:1}; note that these have mostly been obtained in
the SSM) and amount to $M_{V'}>2.5$ TeV and 3.6 TeV in the UU and NU
models, respectively.

%
\begin{table}[h]
 \center
 \begin{tabular}{cccccc}
 \hline\hline
 Name   & Model & $M_{W'}$ [TeV] & $t$ & $\Gamma_{W'}$ [GeV]& $\Gamma_{W'\to\ell\nu}$ [GeV] \\
 \hline\hline
 $B_{1}$ & SSM & 4 & --- & 142.85 & 11.69 \\
 $B_{2}$ & UU  & 4 & 0.7 & 237.15 & 5.73  \\
 $B_{3}$ & UU  & 4 & 1.2 & 125.35 & 16.83 \\
 $B_{4}$ & NU  & 4 & 0.7 & 217.80 & 23.85 \\
 $B_{5}$ & NU  & 4 & 1.4 & 141.82 & 5.96  \\
 \hline\hline
 \end{tabular}
 \caption{Definitions of our SSM and G(221) benchmark points and their
 corresponding total and leptonic decay widths.}
 \label{tab:3}
\end{table}
%
For our benchmark points, listed in Tab.\ \ref{tab:3}, we therefore choose in all
models, including the SSM, a new gauge boson mass of 4 TeV. The allowed ranges in
$|t|$ are then  $[0.18;1.2]$ for the UU and $[0.69;1.47]$ for the NU model. In
these ranges, we select two values of $t$ different from one, which would be similar
to the SSM. While for the upper values we take (almost) maximal choices, the fact that
we limit ourselves for the lower values to 0.7 also in the UU model is due to the
observation that below this value the total decay width of the $W'$ boson becomes very large
and even comparable to its mass.

\subsection{Transverse momentum distributions}

%
\begin{figure}[t]
 \begin{center}
 \includegraphics[width=\textwidth]{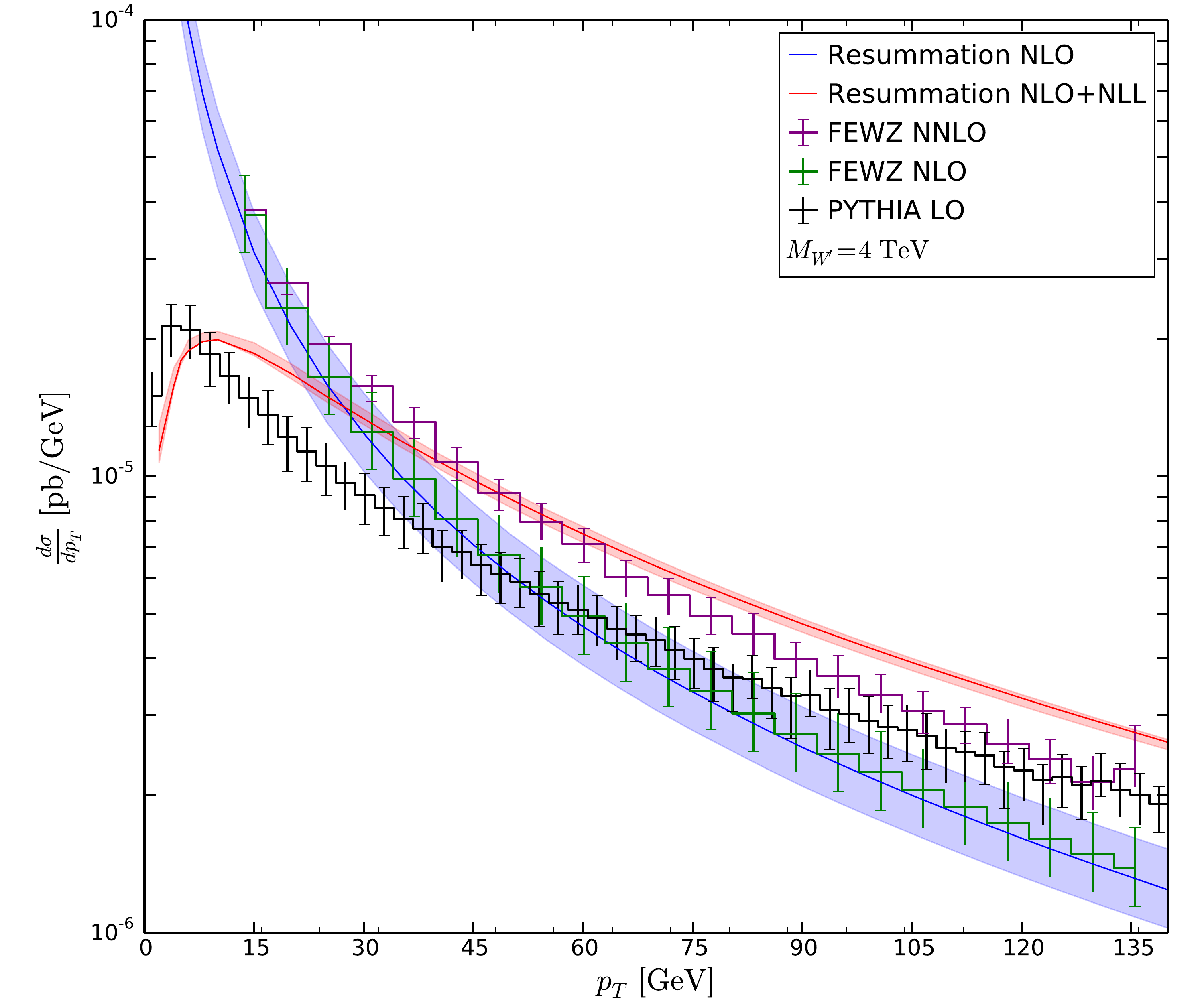}
 \end{center}
 \caption{Transverse momentum distributions of $W'$ bosons with a mass of 4 TeV at the
 LHC14 in the SSM. We compare our NLO and NLO+NLL predictions with RESUMMINO
 to those obtained with PYTHIA LO+PS, FEWZ NLO and NNLO.}
 \label{fig:2}
\end{figure}
%

At LO of perturbative QCD, weak gauge bosons are produced through the Drell-Yan process
with vanishing transverse momentum $p_T$. This changes at NLO (and beyond), when the $p_T$
of the vector boson can be balanced by one (or more) hadronic jet(s). Due to the
incomplete cancellation of soft gluon radiation, the $p_T$ spectrum diverges at fixed
order (see Sec.\ \ref{sec:2}), and only after resummation of the QCD corrections to
all orders a finite spectrum is obtained.

This can be observed in Fig.\ \ref{fig:2} for positively charged $W'$ bosons of mass 4 TeV
produced at LHC14 in the SSM and assumed to decay into a positron and an electron neutrino.
In order to enhance the contribution from the new gauge boson and limit the one from the
SM $W$ boson as well as interference effects, we have implemented a cut on the invariant
mass of the lepton pair of $Q>3M_{W'}/4$. The NLO predictions obtained with FEWZ and
RESUMMINO then agree very nicely, both for their central values and for their scale
uncertainties, and both diverge as $p_T\to0$. In contrast, the LO $\delta$-distribution
(not shown) is modified by the PYTHIA PS to a finite distribution, which exhibits a maximum around
$p_T\sim7$ GeV. A similar turnover, with a maximum at slightly larger values of $p_T\sim10$
GeV, is exhibited by the resummation calculation at NLO+NLL. The difference in shapes
can be attributed to different logarithmic accuracies (LL in PYTHIA, NLL in RESUMMINO),
while the one in normalisation comes mostly from the different perturbative order
(LO in PYTHIA, NLO+NLL in RESUMMINO). At higher $p_T$ values, the NLO+NLL resummation
calculation agrees better with the fixed-order one by FEWZ at NNLO than at NLO, indicating
that important contributions beyond NLO are captured in the resummation approach. Also
the scale errors of these higher-order calculations are then comparable.

\subsection{Total cross sections}

If we integrate (by eye) over the transverse momentum distribution in Fig.\ \ref{fig:2},
we see that in the SSM (and similarly in the UU and NU models) one can expect the total
cross sections for positively charged $W'$ bosons of mass 4 TeV decaying into positrons
and neutrinos to reach about 1 fb at LHC14. This is indeed the case, as one observes in
Tabs.\ \ref{tab:4} and \ref{tab:5} for our five different benchmark points defined
in Tab.\ \ref{tab:3}.

%
\renewcommand{\arraystretch}{1.3}
\begin{table}[t]
 \center
 \resizebox{\columnwidth}{!}{%
 \begin{tabular}{cccccccc}
 \hline\hline
 Model & RESUMMINO LO & PYTHIA LO & RESUMMINO NLO & FEWZ NLO & RESUMMINO NLO+NLL & FEWZ NNLO\\
 \hline\hline
 $B_{1}$& $1338.6^{-155.5}_{+186.7}$	&	$1366.8^{+190.3}_{-158.6}$  &	$1469.2^{+119.7}_{-134.7}$	&	$1492.9^{+74.7}_{-79.4}\pm^{+127.9}_{-89.2}$    	&	$1411.2^{-88.7}_{-37.2}$	&	$1509.1^{+25.7}_{-34.5}\pm^{+146.9}_{-92.3} $  	\\
 $B_{2}$& $799.2^{+92.5}_{-111.4}$	&	$833.2^{+116.3}_{-96.4}  $ 	&$874.6^{+73.8}_{-83.9}$	&	$893.5^{+44.7}_{-47.3}\pm^{+74.9}_{-52.0}$      	&$843.3^{-47.5}_{-26.0}$		&	$902.7^{+12.7}_{-18.4}\pm^{+86.5}_{-54.3}$     	\\	
 $B_{3}$&$ 1515.4^{+175.3}_{-213.6}$		&	$1552.6^{+217.4}_{-179.0}$ 	&$1672.7^{+138.9}_{-156.2}$		&	$1689.2^{+85.5}_{-90.3}\pm^{+145.2}_{-101.4}$   	&$1605.7^{-99.7}_{-44.2}$		&	$1705.1^{+24.2}_{-35.3}\pm^{+168.1}_{-105.7}$   	\\
 $B_{4}$&$3630.9^{+420.3}_{-506.5} $		&	$3669.1^{+512.8}_{-425.9}$ 	&$3986.9^{+339.9}_{-375.4}$		&	$4053.5^{+203.3}_{-215.3}\pm^{+341.0}_{-236.9}$ 	&$3841.5^{-214.4}_{-112.1}$		&	$4094.5^{+57.6}_{-83.7}\pm^{+394.3}_{-247.6}$  	\\
 $B_{5}$&$351.2^{+41.1}_{-49.0}$		&	$383.3^{+53.7}_{-44.5}   $ 	&	$385.2^{+31.3}_{-35.7}$	&	$388.9^{+19.6}_{-20.8}\pm^{+47.8}_{-33.4}$	     	&	$369.9^{-23.4}_{-10.2}$	&	$392.6^{+5.5}_{-8.1}\pm^{+38.5}_{-24.2}$	       	\\
 \hline\hline
 \end{tabular}
 }
 \caption{Total cross section predictions for positively charged $W'$ bosons decaying into a
 positron and a neutrino at LHC14 (in attobarns) for the benchmark points defined in Tab.\
 \ref{tab:3}. Interference terms between $W$ and $W'$ gauge bosons are neglected. The
 invariant mass of the lepton pair is restricted to $Q>3M_{\Wp}/4$.}
 \label{tab:4}
\end{table}
\renewcommand{\arraystretch}{1}
%

%
\renewcommand{\arraystretch}{1.3}
\begin{table}[t]
 \center
 \resizebox{\columnwidth}{!}{%
 \begin{tabular}{cccccc}
 \hline\hline
 Model & PYTHIA w/o int. & PYTHIA w/ int. & RESUMMINO LO& RESUMMINO NLO & RESUMMINO NLO+NLL\\
 \hline\hline
 $B_{1}$	&	$1366.8^{+190.3}_{-158.6}$      &	   $1237.7^{+175.4}_{-145.5}$	   &$ 1241.7^{+147.6}_{-176.1}$	  &	$1379.5^{+113.4}_{-121.1}$		     &	$1313.3^{-92.3}_{-27.9}$	\\		
 $B_{2}$	&	$833.2^{+116.3}_{-96.4}  $	   	&    $953.2^{+128.1}_{-108.6}$     &$949.0^{+107.6}_{-129.5} $   &    $1013.8^{+90.3}_{-105.7}$            &   $993.1^{-37.7}_{-40.0}$     \\
 $B_{3}$	&	$1552.6^{+217.4}_{-179.0}$	   	&    $1684.3^{+234.3}_{-194.4}$    &$1676.9^{+193.5}_{-233.0}$  	 &    $1831.2^{+158.9}_{-177.3}$           &   $1775.6^{-86.7}_{-57.2}$    \\
 $B_{4}$	&	$3669.1^{+512.8}_{-425.9}$	   	&    $3418.0^{+478.2}_{-404.0}$    &$3419.4^{+398.8}_{-481.6}$	    &    $3781.1^{+318.5}_{-343.2}$           &   $3618.7^{-228.5}_{-90.3}$    \\
 $B_{5}$	&	$383.3^{+53.7}_{-44.5}   $	   	&    $317.9^{+45.3}_{-37.8}$       &$317.9^{+37.6}_{-45.8}$    &    $351.9^{+29.5}_{-32.9}$              &   $332.7^{-25.4}_{-9.0}$      \\
 \hline\hline
 \end{tabular}
 }
 \caption{Same as Tab.\ \ref{tab:4}, but with interference terms now included.}
 \label{tab:5}
\end{table}
\renewcommand{\arraystretch}{1}
%

For a more precise comparison, it is first mandatory to remove interference effects
from the PYTHIA and RESUMMINO predictions, as these are not implemented in FEWZ.
Then, the predictions with comparable accuracy in Tab.\ \ref{tab:4} can be seen to
agree within 1-2 percent for their central values and also, although somewhat less
precisely, for their scale errors. First, this is the case for PYTHIA LO, where the
PS does not alter the total cross section, and RESUMMINO LO, both computed with MSTW
2008 LO PDFs. Second, this is also the case for the NLO predictions of RESUMMINO and
FEWZ.\footnote{Unfortunately, our attempts to bring the RESUMMINO and FEWZ NLO
predictions in agreement with those of the $W'$ versions of MC@NLO and POWHEG
\cite{Papaefstathiou:2009sr} failed after replacing there the default squared scales
$\mu_R^2=\mu_F^2=ut/s-Q^2$ with our default choice $M_{W'}^2$ and intensive discussions
with the authors and despite the fact that inferferences seem to be
implemented there correctly.}
Finally, the RESUMMINO NLO+NLL predictions are seen to be stabilised with
respect to their NLO central values and scale errors, while the FEWZ full NNLO
predictions are again somewhat larger. The larger disagreement between RESUMMINO and
FEWZ at this level can be traced to the fact that we are not yet close enough to
the threshold region, where resummation calculations are most reliable. In the last
column, we also give the PDF error computed with FEWZ at NNLO using MSTW 2008 NNLO
error PDFs. As one can see, at this precision these errors largely dominate over
the scale errors, since they are not only sensitive to higher-order corrections,
but also to the experimental errors entering the global fit procedure.

%
\begin{figure}[t]
 \begin{center}
 \includegraphics[width=\textwidth]{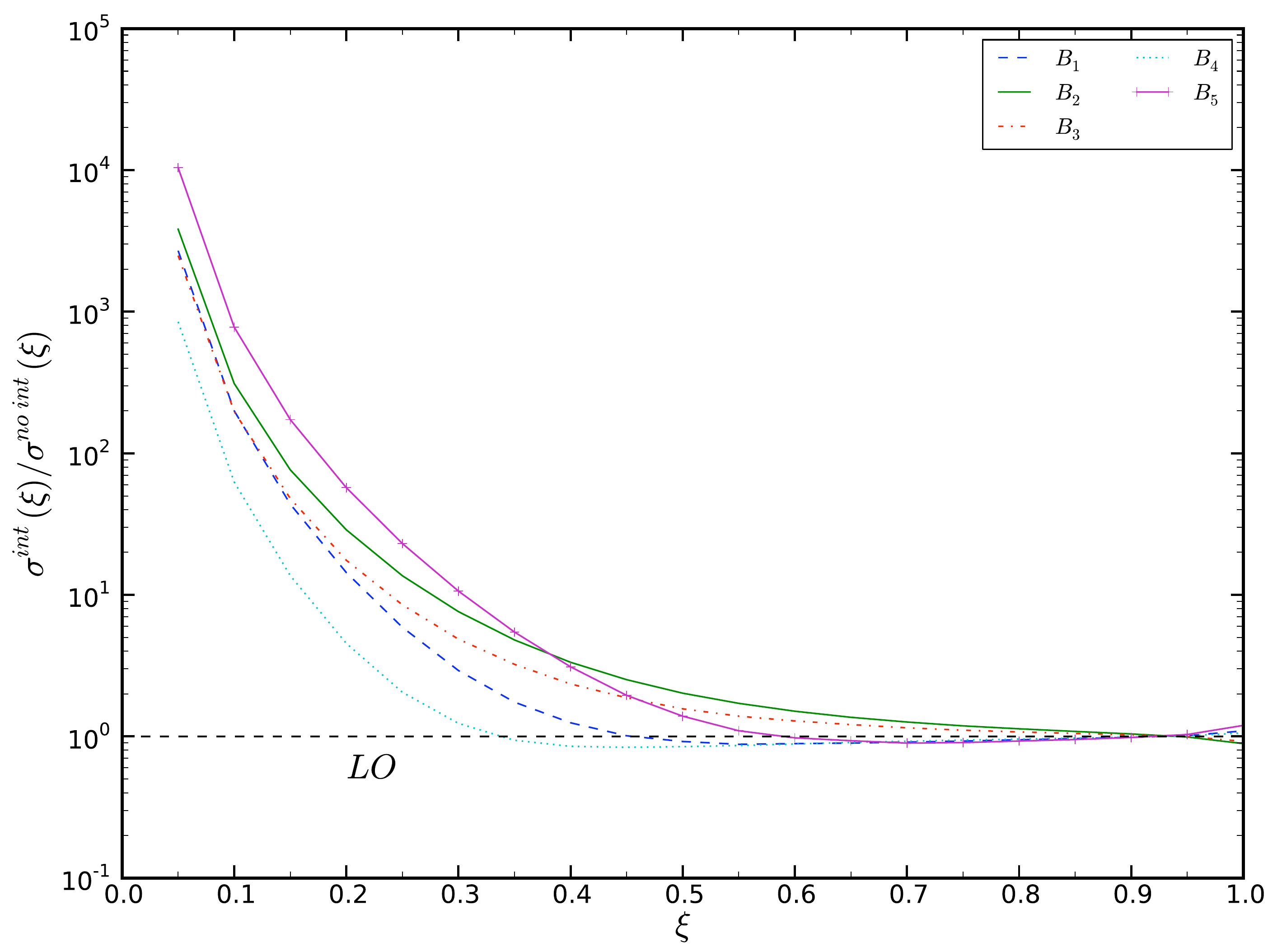}
 \end{center}
 \caption{Ratios of the total cross section at LO with and without interference
 terms as a function of the minimal invariant mass cut $Q>\xi M_{W'}$ for
 our five benchmark points.}
 \label{fig:3}
\end{figure}
%
Looking at Tab.\ \ref{tab:5}, we observe that interference effects can quite
significantly affect the total cross section predictions despite the invariant
mass cut of $Q>3M_{W'}/4$. Depending on the model and benchmark point, the PYTHIA
LO predictions decrease or increase by up to +14\% (for $B_2$) and -17\% (for $B_5$).
When interference effects are also included in RESUMMINO, the agreement with
PYTHIA at LO is nevertheless as good as before. Again, a significant increase
in total cross section at NLO is followed by a stabilisation at NLO+NLL, both
in the central value and in the reduction of the scale error.

Let us investigate somewhat further the effect of the invariant mass cut on
the importance of interference contributions. As one can see in Fig.\ \ref{fig:3},
these become quickly
dominant as the invariant mass cut falls below 50\%. This will become important
in Sec.\ \ref{sec:4}, when we reanalyse the latest ATLAS and CMS results on $W'$ 
and $Z'$ boson
production. But note that even for a cut of 75\% as we employ here, the
interference terms can still modify the total cross section prediction by almost
20\% as we have also observed above. Depending on the model and the applied cut,
the change can be both positive and negative.

%
\begin{figure}[t]
 \begin{center}
 \includegraphics[width=0.9\columnwidth]{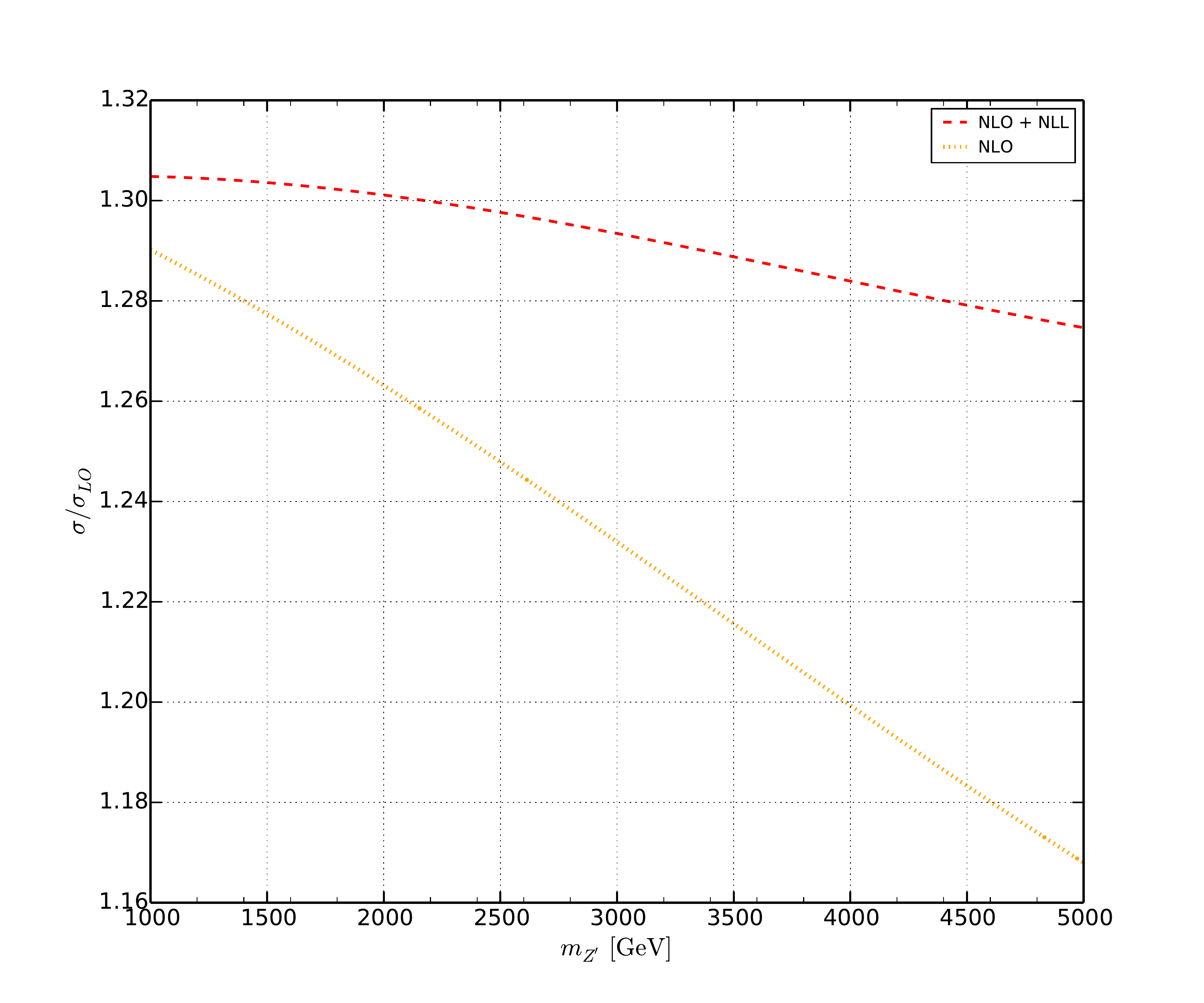}
 \end{center}
 \caption{Ratios of $\Zp$ production cross sections at LHC14 at NLO and NLO+NLL over the
 LO cross section in the SSM and as a function of the heavy gauge boson mass.}
 \label{fig:4}
\end{figure}
%

To end this section, we study in Fig.\ \ref{fig:4} the dependence of the
resummation contributions on the new gauge boson mass, using now the example
of a neutral $Z'$ gauge boson produced at LHC14. Since we show the ratios of
NLO and NLO+NLL cross sections over the LO one, the decay channel is not
relevant. As usual, the NLO QCD corrections to the total cross section are
quite important. For the $Z'$ boson masses considered here, they amount to
29-17\%, i.e.\ seem to decrease with increasing mass. A look at the NLO+NLL
prediction shows that as one approaches the threshold region the resummation
of logarithms becomes increasingly important, i.e.\ the QCD corrections remain
at a similar level of about 28\% even in the high mass region. Therefore our
resummation calculations will become even more relevant as the LHC explores
higher and higher mass regions.

\section{Gauge boson mass limits in general SM extensions}
\label{sec:4}

In this section, we reanalyse the latest experimental searches by the ATLAS and CMS
collaborations for $W'$ and $Z'$ bosons in their leptonic decay channels, performed at LHC8 in
the SSM. We use our resummation predictions at NLO+NLL and do this not only in the SSM,
but also in the UU and NU models that have previously not been considered. 

\subsection{ATLAS limits on $W'$ boson masses}

%
\begin{figure}[b!]
 \centering
 \includegraphics[width=\textwidth]{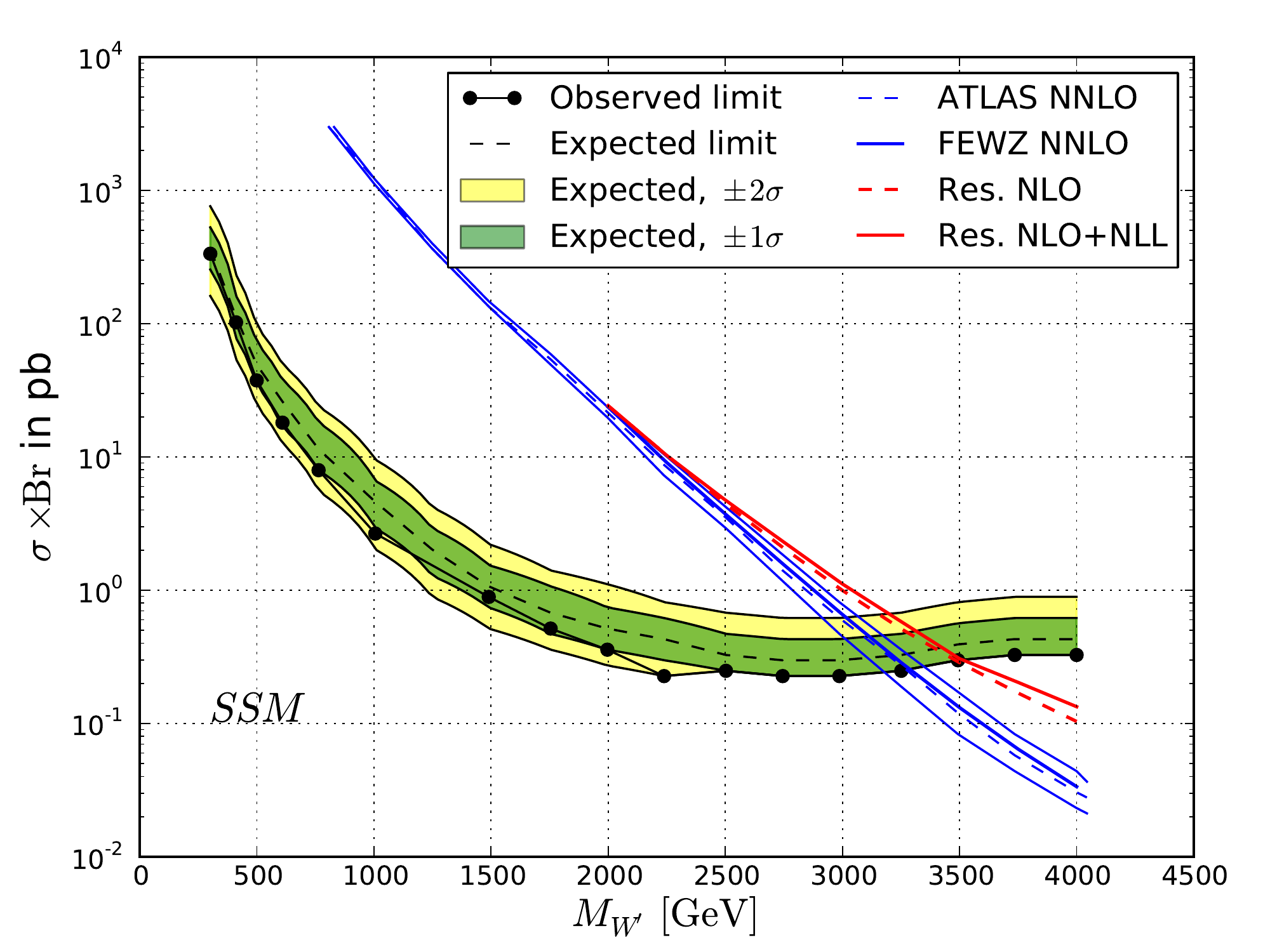}
 \caption{Cross sections times branching ratios for SSM $W'$ bosons decaying into electrons or muons
 and neutrinos at LHC8. The limits expected (dashed black) and observed (full black) in the preliminary ATLAS
 analysis \cite{ATLAS:2014fk}, using a cut of $Q>0.4M_{W'}$ at the generator level, and their corresponding
 uncertainties at the 1$\sigma$ (green) and 2$\sigma$ (yellow) level are compared to predictions
 without interference at NNLO in ZWPROD (with the dominating PDF uncertainties, dashed blue) and in
 FEWZ (central only, full blue) and with interference at NLO (central only, dashed red) and at NLO+NLL (central only,
 full red) using RESUMMINO.}
  \label{fig:5}
\end{figure}
%

The preliminary ATLAS limit of $M_{W'}>3.27$ TeV \cite{ATLAS:2014fk} in the SSM is almost
identical to the corresponding CMS limit of 3.28 TeV \cite{Khachatryan:2014tva}. In their
preliminary analysis, the ATLAS collaboration employ an invariant mass cut of $Q> 0.4 M_{W'}$
at the generator level, which we can directly implement in our theoretical predictions with
RESUMMINO, in contrast to a minimal cut on the missing transverse mass. This minimal cut
is the distinctive variable in the final ATLAS \cite{ATLAS:2014wra} and CMS \cite{Khachatryan:2014tva}
analyses, where the former led to $M_{W'}>3.24$ TeV, i.e.\ again almost identical to the
preliminary ATLAS result. This similarity can be traced to the fact that the invariant mass cut mimicks
very well the other experimental cuts; in particular, practically no signal cross section is lost. We can therefore
be confident that our reanalysis of the preliminary ATLAS results also holds with very good accuracy for
the published ATLAS results.

Both the preliminary and final ATLAS analyses are performed by simulating the $W'$ signal
with PYTHIA LO+PS, adding negative and positive charges, and rescaling it to NNLO total
cross section accuracy with ZWPROD \cite{Hamberg:1990np}. This means, however, that
interference effects between SM $W$ bosons and SSM $W'$ bosons are not included. As one
can see by comparing the original ATLAS NNLO prediction with ZWPROD (dashed blue curve)
to ours with FEWZ (full blue curve) in Fig.\ \ref{fig:5}, they are basically identical, validating our
re-analysis for settings identical to those in the ATLAS analysis. The theoretical error (blue
band),
dominated at NNLO by the PDF uncertainties as parameterised in the MSTW 2008 NNLO
error sets at 68\% C.L., increases with the mass of the $W'$ boson to about $\pm30$\% at
$M_{W'}=4$ TeV. Looking  at the RESUMMINO predictions that include interference effects,
these are seen to be very important, since the invariant mass cut is relatively low (cf.\ Fig.\ \ref{fig:3}),
and they lead to an increase of
$\sigma\times$Br of about a factor of two at the highest mass considered here. There, the
resummation effects are also best visible, and they increase the NLO prediction (dashed red)
by about 20\% at NLO+NLL (full red). Note that these numbers can not be directly compared
to those in Tabs.\ \ref{tab:4} and \ref{tab:5}, where the invariant mass cut was $Q>3M_{W'}/4$.
In the SSM, we can then exclude $W'$ bosons with masses below 3.5 TeV.

%
\begin{figure}[t]
 \centering
 \includegraphics[width=\textwidth]{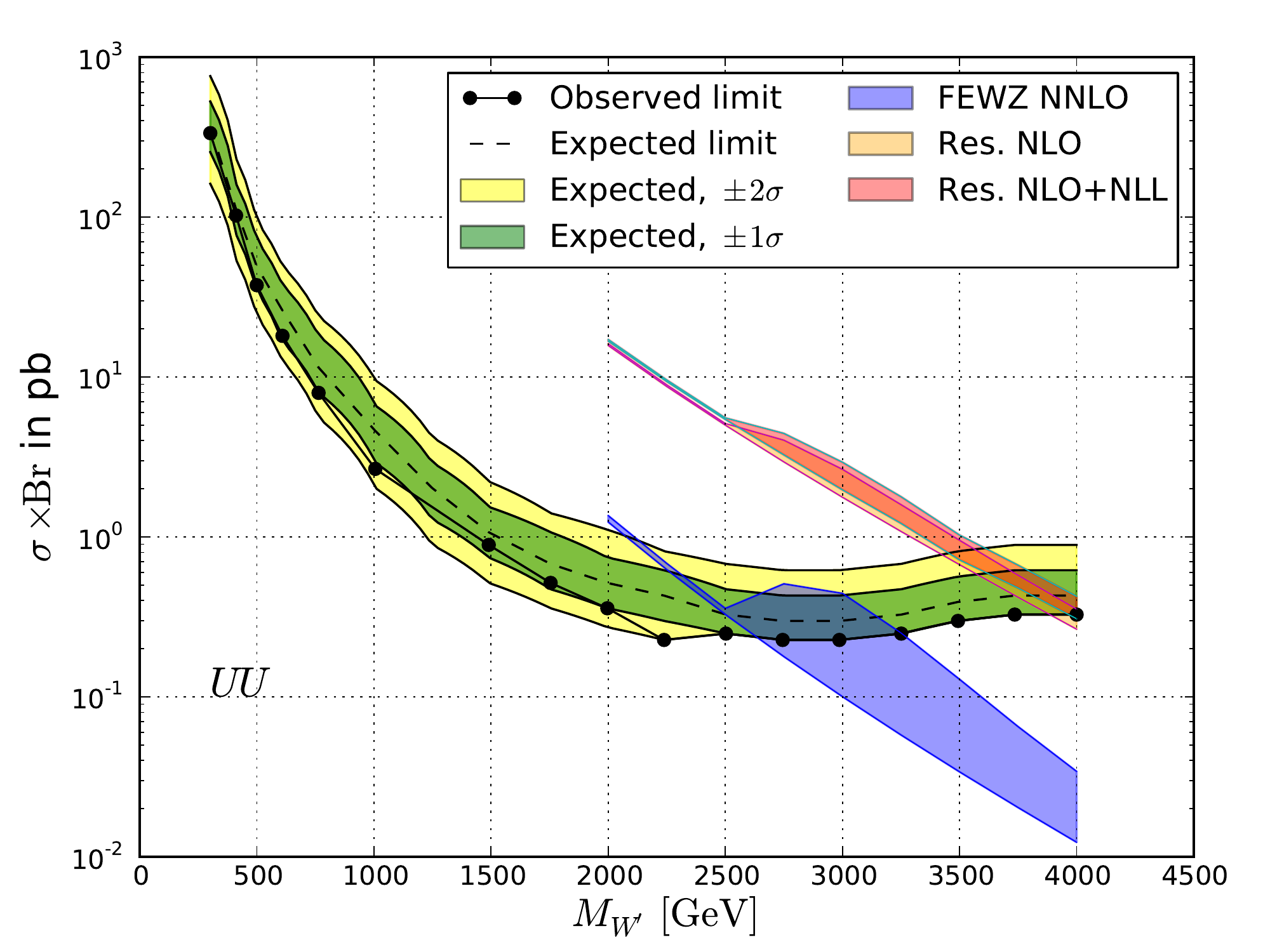}
 \caption{Same as Fig.\ \ref{fig:5} for UU model $W'$ bosons.}
  \label{fig:6}
\end{figure}
%

The results for the UU model are presented in Fig.\ \ref{fig:6}. There, the expected and observed
ATLAS limits are first compared to FEWZ NNLO (blue) 
results without interference. For each $W'$ boson mass, the variation of the $t$ parameter
in the allowed range (see Fig.\ \ref{fig:1}) leads to a spread of theoretical predictions. This is reflected
in the shown areas, which basically overlap for FEWZ NNLO and RESUMMINO NLO+NLL (not shown). Note, 
however, that part of these areas correspond to values of $t$ below 0.7, where the $W'$ width in the
UU model becomes very large. The inclusion of interference effects in RESUMMINO leads
to an increase of the predicted cross sections by almost an order of magnitude at $M_{W'}=4$ TeV.
There, the predictions at NLO (light red) are increased by less than  20\% at NLO+NLL (dark
red).  Below masses of 2.5 TeV, where the UU model is already excluded by low-energy and
precision constraints \cite{Hsieh:2010zr}, the areas have been shrunk to a single line, calculated
for a hypothetical $t$-value of 0.18 pertinent at the same time to the minimal allowed mass and the
perturbativity limit. At NLO+NLL and including interference effects, our reanalysis excludes $W'$ bosons in
the UU model with masses below 3.9--4 TeV, which considerably improves the limits from low-energy
and precision constraints. As in this model $M_{Z'}\simeq M_{W'}$ up to corrections of ${\cal O}
(v^2/u^2)$, this implies an identical mass limit for $Z'$ bosons in the UU model.

%
\begin{figure}[t]
 \centering
 \includegraphics[width=\textwidth]{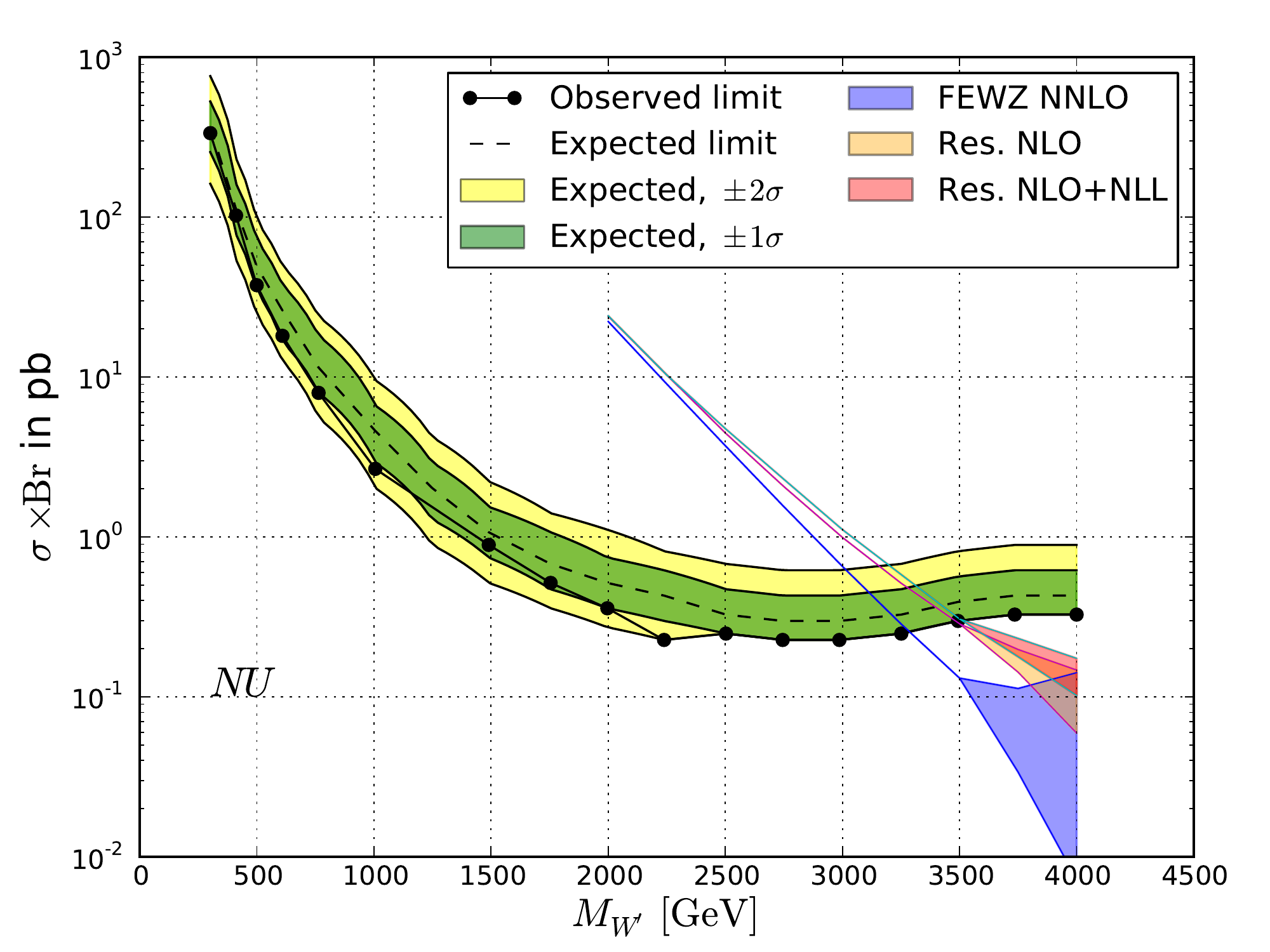}
 \caption{Same as Fig.\ \ref{fig:5} for NU model $W'$ bosons.}
  \label{fig:7}
\end{figure}
%
Our analysis in the NU model is shown in Fig.\ \ref{fig:7}. Without interference, the FEWZ NNLO
(blue) and RESUMMINO NLO+NLL (not shown) results agree again, i.e.\ the regions spanned
by the allowed $t$ values above the minimal mass of 3.6 TeV overlap. Inteference effects increase
the predicted cross sections by about a factor of two in the high mass region, while the NLO+NLL
results (dark red) are about 20\% larger than the NLO results (light red), both computed with
RESUMMINO. In this case, the ATLAS data do not improve on the low-energy and precision
constraints, but only lead to a slightly weaker exclusion bound of $W'$ bosons in the NU model
of about 3.5 TeV. As above, the same limit applies also to $Z'$ boson masses in the NU model.

\subsection{CMS limits on $Z'$ boson masses}

The CMS collaboration have searched for narrow resonances in the dilepton
(electron or muon) mass spectrum and set mass limits of 2.96 TeV and 2.6 TeV
on SSM $Z'$ bosons and a specific class of superstring-inspired $Z'$ bosons,
respectively \cite{CMS:2013qca}. The final ATLAS SSM limit of 2.90 TeV is only
slightly weaker \cite{Aad:2014cka}. While the ATLAS collaboration set limits
directly on the new gauge boson production cross section times branching ratio,
$\sigma\times$Br, the CMS collaboration set limits on the ratio $R_\sigma$ of 
this quantity for the $Z'$-boson to the one for the SM $Z$-Boson.

\begin{figure}[t]
  \centering
  \includegraphics[width=\textwidth]{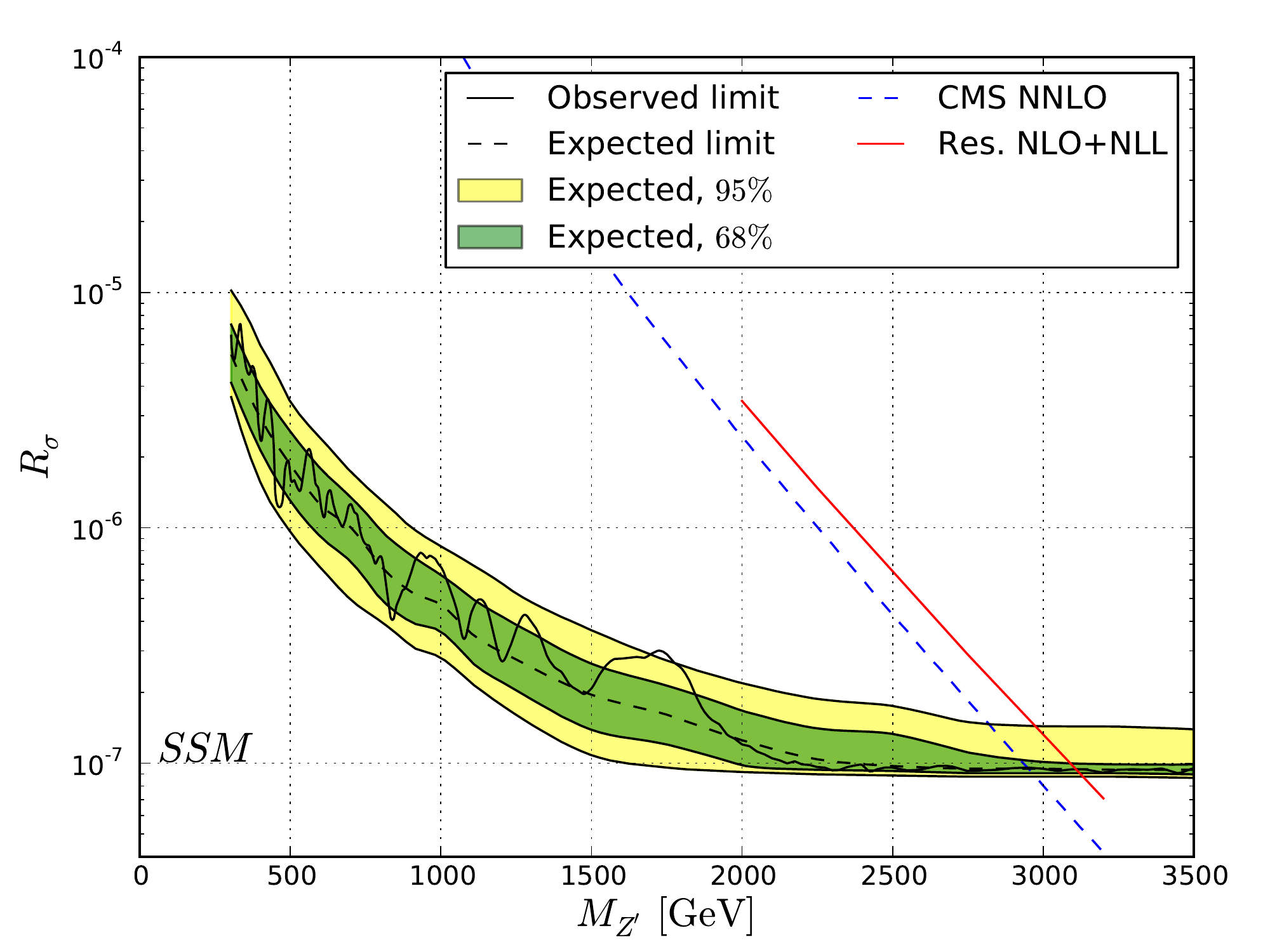}
  \caption{Ratios of new physics over SM cross sections for SSM $Z'$ bosons decaying into electron or muon
 pairs at LHC8. The limits expected (dashed black) and observed (full black) in the final CMS
 analysis \cite{CMS:2013qca}, using a cut of $0.6M_{Z'}<Q<1.4M_{Z'}$, and their corresponding
 uncertainties at the 68\% (green) and 95\% (yellow) C.L.\ are compared to predictions
 without photon, $Z$ and $Z'$ interference at NNLO in ZWPROD (dashed blue)
 and with full interference at NLO+NLL (full red) using RESUMMINO.}
  \label{fig:8}
\end{figure}

The mass limits are obtained by comparing expected and observed experimental
limits on $R_\sigma$ with expectations from PYTHIA LO+PS, rescaled to NNLO
with ZWPROD. For the SSM, we show the result in Fig.\ \ref{fig:8}, where one
%
%
can read off the limit cited above. While interferences between the $Z'$ boson and SM
contributions are not included in the numerator, those of SM $Z$ bosons and photons have
been included in the denominator of this ratio, as we have verified by comparing with
FEWZ at NNLO. Adding the interferences also in the numerator leads to a considerable
increase of the prediction, computed by us with RESUMMINO at NLO+NLL, so that the SSM
exclusion limit moves to 3.2 TeV.

%
\begin{figure}[t]
  \centering
  \includegraphics[width=\textwidth]{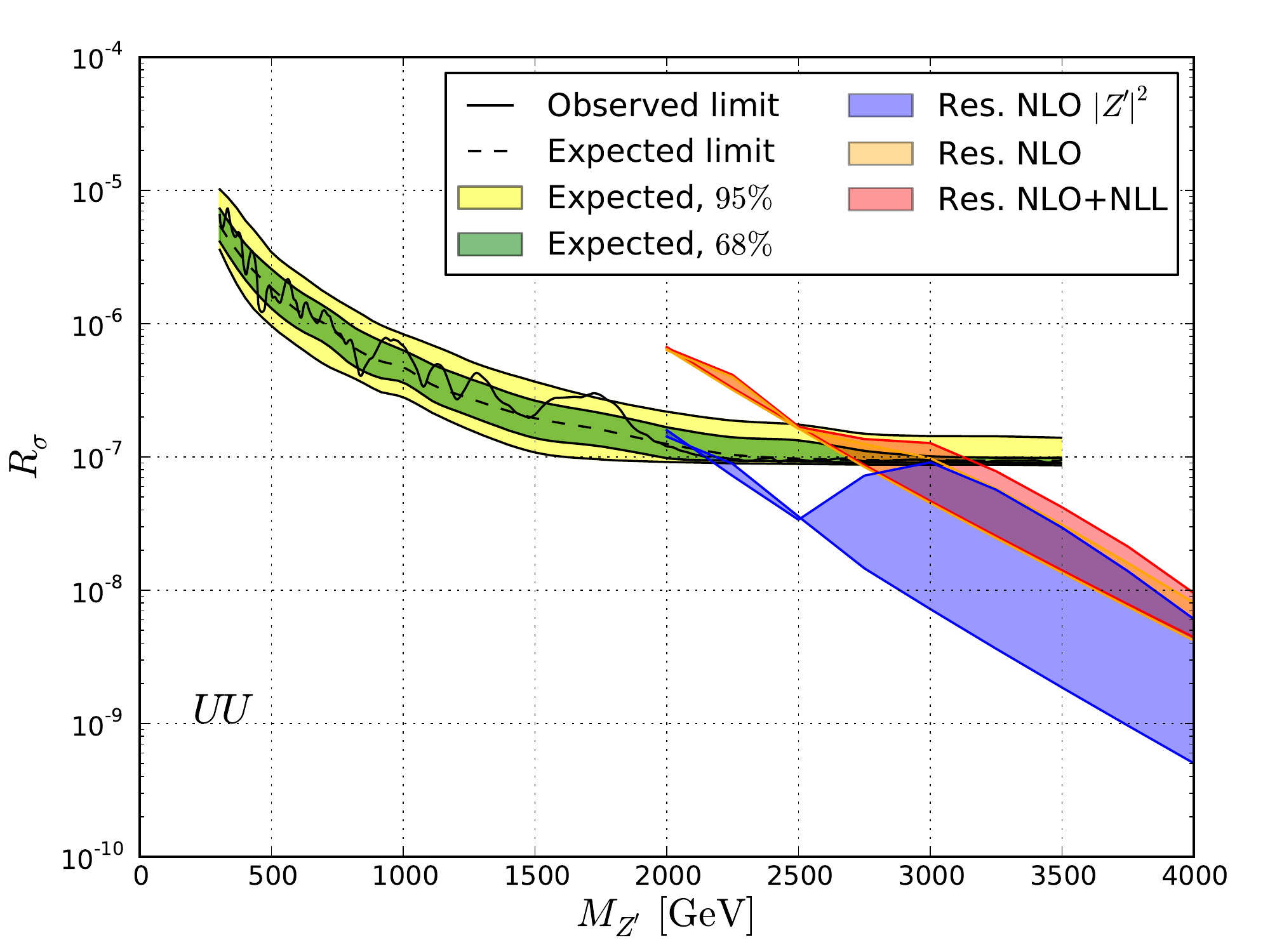}
  \caption{Same as Fig.\ \ref{fig:8} for UU model $Z'$ bosons.}
  \label{fig:9}
\end{figure}
%
For $Z'$ bosons in the UU model, we simulate in Fig.\ \ref{fig:9}
the ratio $R_\sigma$ without interference in the numerator using RESUMMINO at NLO accuracy
(blue area). Interference effects then increase again the prediction (light red) by
about an order of magnitude, while the additional radiative corrections at NLO+NLL
(dark red) do not alter the result significantly. This is very likely due to the fact that
these corrections affect both the numerator and the denominator in a similar
way. In the UU model, we then obtain $Z'$ boson mass limits ranging from 2.75 TeV up to
3.2 TeV, depending on the chosen value of the parameter $t$. These are in all cases
stronger than the previously obtained indirect limit of 2.5 TeV.

\begin{figure}[t]
  \centering
  \includegraphics[width=\textwidth]{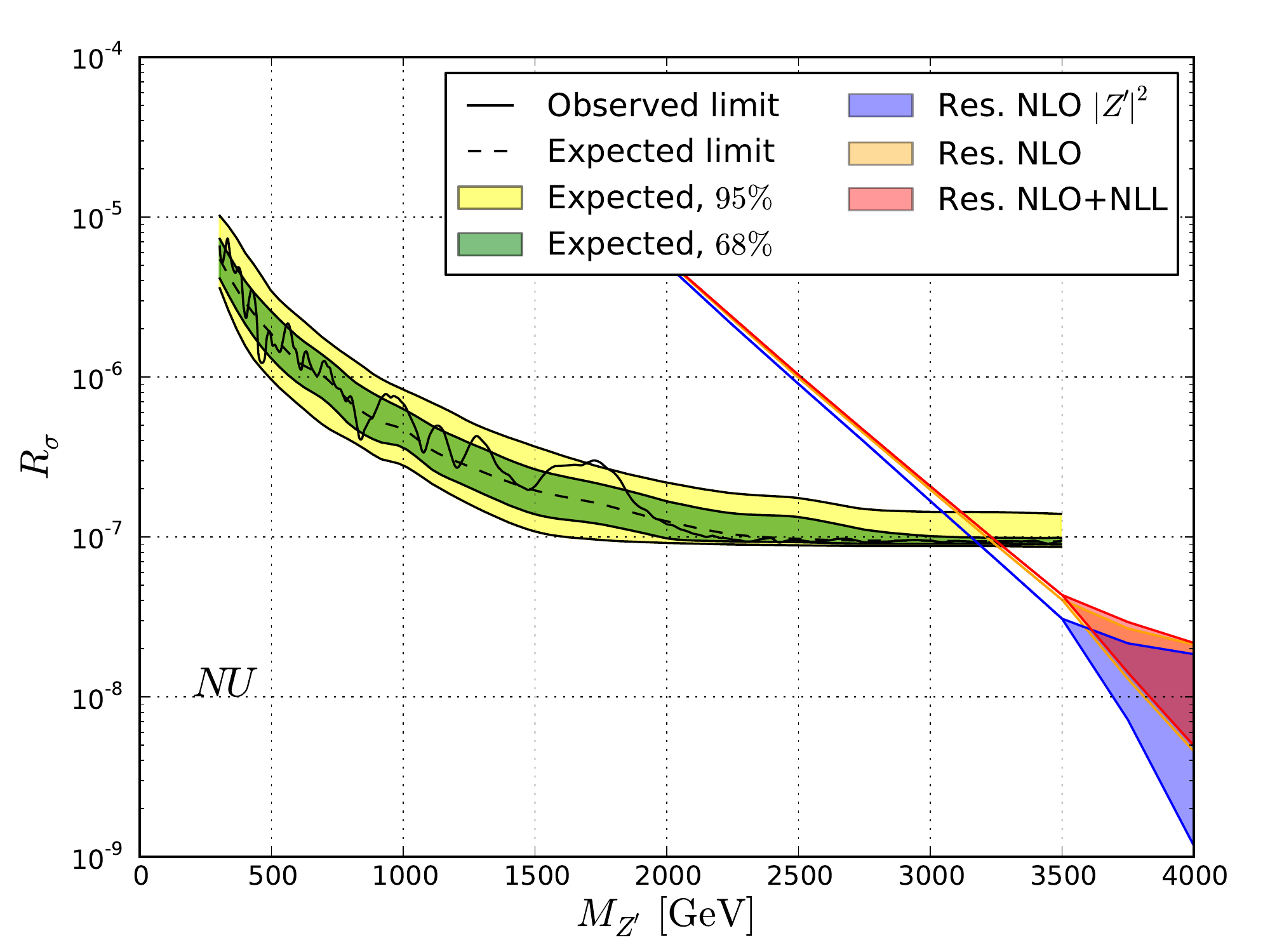}
  \caption{Same as Fig.\ \ref{fig:8} for NU model $Z'$ bosons.}
  \label{fig:10}
\end{figure}

For NU model $Z'$ bosons, shown in Fig.\ \ref{fig:10}, the interference effect is somewhat
less pronounced, but still clearly visible, while radiative effects are again relatively
small in the ratio $R_\sigma$. Similarly to our reanalysis of the ATLAS $W'$ search, we can
only set a lower mass limit of 3.25 TeV, which does not exceed the one of 3.6 TeV
obtained from precision measurements and at lower energy.

%
%

\section{Conclusion}
\label{sec:5}

In this paper, we have presented resummation calculations at NLO+NLL accuracy for the
production of leptonically decaying $W'$ and $Z'$ bosons in hadronic collisions at small
transverse momenta and/or close to production threshold. Our calculations include the full
interference structure of new and SM gauge bosons, which is unfortunately missing from
full NNLO calculations. They therefore currently provide the best available theoretical precision
for realistic cross section estimates. To facilitate a comparison with LO+PS calculations, we
furthermore implemented interference effects in PYTHIA by adding a new $2\to2$ process,
i.e.\ without relying on resonant production or the narrow width approximation. 

We demonstrated that in the SSM the PYTHIA transverse momentum spectrum of $W'$ bosons
with a mass of 4 TeV agrees qualitatively with our resummation calculations at low $p_T$,
whereas at intermediate $p_T$ the resummed predictions lie close to those at NNLO, showing
that a substantial fraction of higher-order corrections is captured by the resummation procedure.

The total cross sections were shown to be stabilised at NLO+NLL compared to the NLO predictions
with respect to variations of the renormalisation and factorisation scales, so that the theoretical
error became dominated for large masses by the PDF uncertainties. Full agreement could be
found at LO with PYTHIA and at NLO with FEWZ -- albeit only without interference. The 
interference effects were shown to depend strongly on the minimal cut on the invariant mass
of the lepton pair, and the resummation contributions were shown to become increasingly
important with the new gauge boson mass.

We did not restrict our analysis to the SSM, but generalised it to G(221) models with an
extended gauge group that could be realised at intermediate scales.
In particular, through a reanalysis of the currently strongest
ATLAS exclusion limits of $W'$ boson masses in the SSM, we showed that $W'$ boson masses could
be excluded below 3.9--4 TeV in the UU model, while the limit of 3.5 TeV in the NU model turned
out to be slightly weaker than the existing low-energy and precision limit of 3.6 TeV. Similarly,
a reanalysis
of the currently strongest CMS exclusion limits of $Z'$ boson masses in the SSM led to
exclusion limits of 2.75--3.2 TeV and 3.25 TeV, which were again stronger in the UU model
and slightly weaker
in the NU model than the low-energy and precision constraints of 2.5 and 3.6 TeV, respectively.
For convenience, our final results in the SSM, UU and NU models for the old and our new
$W'$ and $Z'$ boson mass limits have been collected in Tab.\ \ref{tab:6}.

%
\begin{table}[t]
 \center
 \begin{tabular}{lcrr}
 \hline\hline
 Model & New gauge boson & Previous mass limit [TeV] & New mass limit [TeV] \\
 \hline\hline
 SSM   & $W'$ & 3.27--3.28 & 3.5 \\
 SSM   & $Z'$ & 2.90--2.96 & 3.2 \\
 UU    & $W'$ & 2.48       & 3.9--4.0 \\
 UU    & $Z'$ & 2.48       & 2.8--3.2 \\
 NU    & $W'$ & 3.56       & (3.5) \\
 NU    & $Z'$ & 3.56       & (3.3) \\
 \hline\hline
 \end{tabular}
 \caption{Previously obtained exclusion limits, using ATLAS \cite{ATLAS:2014fk,Aad:2014cka}
 and CMS data \cite{Khachatryan:2014tva,CMS:2013qca} for the SSM as well as low-energy and
 precision data for the UU and NU models \cite{Hsieh:2010zr}, and new exclusion
 limits, including all interference effects and NLO+NLL corrections, for $W'$ and $Z'$
 gauge bosons.}
 \label{tab:6}
\end{table}
%

\section*{Note added}
While in general not much attention has been paid to G(221) models, the NU
model has recently been studied in Ref.\ \cite{Kim:2014afa}. Simulations with
standard PYTHIA6.4 (i.e.\ at LO+PS and without interferences) of $\ell\ell$, $jj$, $\tau\tau$
and $tt$ final states for $Z'$-bosons and of $\ell\nu$ for $W'$-bosons have been
compared to ATLAS and CMS data, resulting in mass limits of 2 TeV in both cases.
These are considerably weaker than our limits, which turned out to be almost
as strong as those obtained from low-energy and precision measurements.

\section*{Acknowledgment} 
We thank A.\ Papaefstathiou for his efforts to bring the $W'$ implementations in MC@NLO and POWHEG
in agreement with our results. This work has been supported by the BMBF Theorie-Verbund, by
a Ph.D.\ fellowship of the French Ministry for Education and
Research, and by the Theory-LHC-France initiative of the CNRS/IN2P3.


\clearpage

\bibliographystyle{JHEP}
\bibliography{paper}

\end{document}